\documentclass[aps,pra,superscriptaddress,nofootinbib,twocolumn]{revtex4-2}

\usepackage{amsmath,amsthm,amsfonts,amssymb,amscd} 
\usepackage[utf8]{inputenc}
\usepackage{indentfirst}
\usepackage{graphicx}
\usepackage{color}
\usepackage[T1]{fontenc}
\usepackage[american,ngerman,greek,brazilian,british]{babel}
\usepackage{float}
\usepackage[dvipsnames]{xcolor}
\usepackage[colorlinks=true,citecolor=Mulberry,linkcolor=prettygreen,urlcolor=MidnightBlue,hyperindex,breaklinks]{hyperref}
\usepackage{breakurl}
\usepackage{cleveref}
\usepackage{mathtools,xfrac}
\usepackage{bm}
\usepackage{ulem}
\usepackage{enumerate}
\usepackage{sidecap}
\usepackage{microtype}
\usepackage{enumitem}
\usepackage{bbold}
\usepackage{listings}
\usepackage{algorithm}
\usepackage{algpseudocode}
\usepackage{multirow}
\usepackage{setspace}
\usepackage{physics}
\usepackage{dsfont}
\usepackage{soul}

\newcommand{\avg}[1]{\langle#1\rangle}
\newcommand{\id}{\mathbb{1}}

\newcommand{\Lcal}{\mathcal{L}}
\newcommand{\Hcal}{\mathcal{H}}

\newcommand{\Ecal}{\mathcal{E}}

\definecolor{awesome}{rgb}{0.93, 0.53, 0.18}
\definecolor{prettygreen}{RGB}{5,125,143}

\newcommand{\cS}{\mathcal{S}}
\newcommand{\ot}{\otimes}

\newtheorem*{theorem*}{Theorem}

\begin{document}

\title{Designing optimal protocols in Bayesian quantum parameter estimation \\ with higher-order operations}

\author{Jessica Bavaresco}
\email{jessica.bavaresco@unige.ch}
\address{Department of Applied Physics, University of Geneva, Geneva, Switzerland}

\author{Patryk Lipka-Bartosik}
\email{patryk.lipka.bartosik@gmail.com}
\address{Department of Applied Physics, University of Geneva, Geneva, Switzerland}

\author{Pavel Sekatski}
\email{pavel.sekatski@gmail.com}
\address{Department of Applied Physics, University of Geneva, Geneva, Switzerland}

\author{Mohammad Mehboudi}
\email{mohammad.mehboudi@tuwien.ac.at}
\address{Technische Universität Wien, 1020 Vienna, Austria}

\begin{abstract} 
Using quantum systems as sensors or probes has been shown to greatly improve the precision of parameter estimation by exploiting unique quantum features such as entanglement. A major task in quantum sensing is to design the optimal protocol, i.e., the most precise one. It has been solved for some specific instances of the problem, but in general even numerical methods are not known. Here, we focus on the single-shot Bayesian setting, where the goal is to find the optimal initial state of the probe (which can be entangled with an auxiliary system), the optimal measurement, and the optimal estimator function.  We leverage the formalism of higher-order operations to develop a method based on semidefinite programming that finds a protocol that is close to the optimal one with arbitrary precision. Crucially, our method is not restricted to any specific quantum evolution, cost function or prior distribution, and thus can be applied to any estimation problem.
Moreover, it can be applied to both single or multiparameter estimation tasks. We demonstrate our method with three examples, consisting of unitary phase estimation, thermometry in a bosonic bath, and multiparameter estimation of an SU(2) transformation. Exploiting our methods, we extend several results from the literature. For example, in the thermometry case, we find the optimal protocol at any finite time and quantify the usefulness of entanglement.
\end{abstract}

\maketitle

\section{Introduction}

Quantum parameter estimation, also known as quantum metrology or quantum sensing, is at the heart of quantum technologies~\cite{Acin_2018}. The quantitative assessment of some properties of a system, such as magnetic field amplitude, length, temperature or chemical potential, to name a few, is a key task for science and industry. A sensor is a device which manipulates probes interacting with the system of interest in order to readout its properties. Loosely speaking, the sensing becomes quantum, whenever the manipulation of the probes and their interaction with the measured system is governed by quantum physics. Quantum metrology has been very successful in advancing technological frontiers as showcased in several experiments, namely, the detection of gravitational waves~\cite{PhysRevLett.116.061102,PhysRevX.11.021053}, thermometry~\cite{Kucsko2013,PhysRevX.10.011018}, magnetometry~\cite{Budker2007,PhysRevLett.104.133601}, and phase estimation in optical platforms~\cite{Mitchell2004}.

The theory of quantum metrology aims at developing protocols that use optimally the probes and other \textit{metrological resources}---such as quantum correlations, coherence and measurement time---in order to estimate the parameter with minimal error~\cite{Giovannetti_2011,RevModPhys.89.035002,Toth_2014,PARIS_2009,PhysRevLett.96.010401}, and uncovers ultimate limits on the achievable estimation precision~\cite{fujiwara2008fibre, escher2011general, demkowicz2012elusive, sekatski2017quantum, demkowicz2017adaptive,zhou2018achieving}. These limits are usually expressed as bound on the Fisher information (matrix) that must hold in a certain context and are related to the mean squared error (MSE) via a Cram\'er-Rao type bound~\cite{Cramer:107581,Rao1992,gill1995applications,PhysRevLett.72.3439,helstrom1969quantum,PhysRevLett.72.3439,e20090628}. In the single-parameter case, such bounds are often saturable in the regime where the protocol is repeated many times~\cite{kay1993fundamentals,kolodynski2014precision}.

\begin{figure}
    \centering
    \includegraphics[width=\columnwidth]{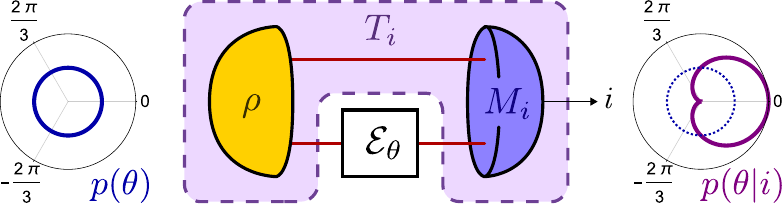}
    \caption{\textbf{Strategy for Bayesian parameter estimation.} The left panel represents the prior probability distribution of the parameter $\theta$ encoded in the channel $\Ecal_\theta$. The center panel shows a single-shot strategy of parameter estimation in which part of a quantum state $\rho$ is sent through the channel $\Ecal_\theta$ and then measured by POVM $\{M_i\}$, yielding a classical outcome $i$. The right panel then represents the posterior probability distribution of the parameter $\theta$, conditioned on the obtained measurement outcome $i$.}
    \label{fig:setup}
\end{figure}

However, in the limit of small data, such bounds are not generally saturable, and furthermore the MSE, addressed  by the Carm\'er-Rao bound, may not be the best quantifier of the estimation precision. Such problems can be attacked from the perspective of the full Bayesian framework. In the Bayesian approach, one starts with a prior distribution (belief) of the parameter and updates it through the protocol based on the observed measurements results. Crucially, the choices of prior distribution and the cost (or reward) function have a substantial impact on the optimal protocol. Finding such optimal Bayesian protocols, is one of the key problems in metrology. This is a non-trivial task even in the case of single-shot scenarios, where the protocol is described by the combination of the initial state, the final measurement, and the estimator function. Optimal protocols are only known for a few highly-symmetric specific cases (see Ref.~\cite{bartlett2007reference,demkowicz2020} for a review),  and for specific cost functions in the single-parameter regime \cite{personick,demkowicz2011optimal}, while general effective numerical methods for finding them are lacking. 

We therefore dedicate this work to address the shortcomings of quantum metrology within the single-shot Bayesian framework. Namely, we exploit the formalism of higher-order operations~\cite{chiribella07,chiribella08transforming,chiribella09,milz23} to combine two pivotal aspects of the estimation protocol---the quantum state and the measurement, referred to as the quantum strategy---into a single and equivalent higher-order transformation, called quantum tester~\cite{chiribella07,chiribella08memory,ziman08}.  While the standard approach to metrology typically involves the optimization over state and measurement individually~\cite{PhysRevResearch.4.043057, kaubruegger23,meyer2023quantum}, often in a non-efficient, heuristic manner, quantum testers allow us to optimize over the quantum strategy altogether, finding  the optimal state and measurement efficiently with a single instance of a semidefinite program (SDP). 
Originally a tool applied to tasks such as channel discrimination~\cite{chiribella08memory,bavaresco21,bavaresco22}, the higher-order operations formalism was recently extended to quantum parameter estimation problem, both in the frequentist setting in order to maximize the Fisher information of a protocol ~\cite{altherr2021,liu2023} and in the Bayesian setting in order to maximize the probability of a fixed-width credible interval~\cite{meyer2023quantum}. In this work we focus on single-shot Bayesian setting and show how to leverage the properties of higher-order operations in order to efficiently optimize the estimation protocol with respect to any reward function.

We propose three different methods to integrate the optimization of the quantum strategy, i.e., state and measurement, with the optimization of the estimators---therefore finding the optimal overall protocol within arbitrary precision. Our methods take into account both numerical and practical limitations, finding application in a wide range of realistic scenarios. It is furthermore appropriate to any estimation problem regardless of prior distribution, reward or cost function, or the type of quantum evolution. Moreover, we show how these methods can be straightforwardly adapted to multiparameter estimation problems.

To demonstrate the merit of our approach, we present three case studies where we apply our methods to relevant parameter estimation problems: phase estimation, thermometry, and SU(2) estimation. These examples cover single and multiparameter problems, both unitary and non-unitary evolution, reward or cost functions of varying nature (e.g. fidelity and MSE), and different prior distributions (e.g. uniform and Gaussian). Moreover, we use one of our case studies, thermometry, to show how our approach can be adapted to approximate quantum strategies that do not permit for entanglement between the probe and an auxiliary system for their implementation.   This allows us to demonstrate that entanglement provides an advantage over no-entanglement strategies in a finite-time temperature estimation task. Our techniques can be similarly used to answer whether entanglement can be useful in other estimation tasks, and put a lower bound on the usefulness of entanglement. In the thermometry problem, we also find the optimal protocol in finite time, which was previously only known in the frequentest regime~\cite{Sekatski2022optimal}, and show that the estimation precision only decreases with $t\to\infty$.

All the code developed for this work is made available in our open online repository~\cite{github-metrology}.

\section{Background}

\subsection{Bayesian parameter estimation}

In a standard metrology problem, one is interested in estimating an unknown parameter $\theta$ by encoding it into the quantum state of a probe. The encoding process can be described by a quantum channel---a completely positive and trace preserving map---which we denote by $\Ecal_{\theta}: \Lcal(\Hcal^I) \to \Lcal(\Hcal^O)$ where $\Hcal^I$ and $\Hcal^O$ are the Hilbert spaces of the input and output  systems of the channel, respectively. When probing the channel, it is in general more advantageous to also use an auxiliary system which is initially entangled to the probe---but does not go through the channel, as sketched in Fig.~\ref{fig:setup}. In other words, one considers the extended channel $\Ecal_{\theta}\otimes \text{id}$, where ``id'' is the identity channel acting on the auxiliary system. The chosen global input state, given by the density operator $\rho \in \Lcal(\Hcal^I\otimes\Hcal^\text{aux})$, is then mapped to a global output state $\rho_{\theta}\coloneqq (\Ecal_{\theta}\otimes\text{id})[\rho]$ by the extended channel. In order to extract the information about the parameter $\theta$ encoded in this state, one performs a joint measurement $M=\{M_i\}_{i=1}^{N_O}$, $M_i\in\Lcal(\Hcal^\text{aux}\otimes\Hcal^O)$ in the auxiliary system and the output state of the channel. Finally, in the considered setting, one designs an \textit{estimator} $\hat \theta$ that assigns an estimate ${\hat{\theta}_i}$ to the true value of the parameter $\theta$, conditioned on each measurement outcome $i$.  The quality of the estimation can be then quantified by setting some score (cost) function, evaluating the closeness (deviation) of the estimator to the true parameter value. Indeed, the score should depend on the \textit{protocol}; i.e., the triplet of the initial state, the measurement, and the estimator $\{\rho,~\{M_i\}_i,~\{\hat{\theta}_i\}_i\}$. A central problem in quantum metrology is finding the optimal protocol.

In the Bayesian approach,  one starts with a prior belief in the parameter value given by a probability distribution $p(\theta)$. After the measurement, described by the Born rule 
\begin{equation}\label{eq::prob_map}
    p(i|\theta) = \tr\left(\rho_{\theta}\,M_i\right) = \tr\left((\Ecal_{\theta}\otimes\text{id})[\rho]\,M_i\right),
\end{equation}
one uses the Bayes' rule to update the distribution of the parameter based on the observed outcome $i$
\begin{align}\label{eq:Bayes}
    p(\theta|i) = \frac{p(i|\theta)p(\theta)}{p(i)},
\end{align}
where the normalization factor is defined as $p(i)\coloneqq \int d{\theta} p(i|\theta)p(\theta)$.

The performance of the estimation strategy can be quantified according to a \textit{score}. Generally, this can be cast as
\begin{align} 
    \cS \coloneqq &\sum_i p(i) \!\int \!d\theta \,p(\theta|i) \, r(\theta,{\hat{\theta}}_i)\\
    = &\int\!d\theta \sum_i \, p(\theta) r(\theta,{\hat{\theta}}_i) \,    \tr\left((\Ecal_{\theta}\otimes\text{id})[\rho]\,M_i\right),   \label{eq::continuous_score}
\end{align}
where $r(\theta,{\hat{\theta}}_i)$ is a reward or cost function that quantifies the difference between the parameter $\theta$ and each estimate ${\hat{\theta}}_i$. A particular choice of cost function is the MSE $r_\text{MSE}(\theta,\hat{\theta}_i) =(\theta - \hat{\theta}_i)^2$. In light of this definition, it becomes clear that the optimal protocol will be the one that either maximizes or minimizes the score ${\cS}$, depending on whether $r(\theta,{\hat{\theta}}_i)$ is a reward or a cost function, respectively. 

As previously mentioned, this problem does not have a known analytical solution in general. and efficient numerical methods have only been proposed for a few special problems~\cite{macieszczak2014bayesian,kaubruegger23}. In this work we provide an efficient algorithm that approximates the solution with arbitrary precision, and works for all cost functions and number of parameters. 

\subsection{Quantum testers: the quantum strategy as a higher-order operation}\label{sec_testers}

A typically cumbersome part of metrology and estimation problems is the optimization of the \textit{quantum strategy}, i.e., of the state and measurement that is used to probe the channel that encodes the parameter to be estimated. Here, we apply techniques from the formalism of higher-order operations~\cite{chiribella07,chiribella08transforming,chiribella09,milz23} to fully characterize the set of quantum strategies applicable to a given estimation task. We then use this reformulation to efficiently optimize over quantum strategies using semidefinite programming~\cite{vandenberghe1996semidefinite,skrzypczyk2023semidefinite,wolkowicz2012handbook}. In particular, we exploit the connection between the states and measurements and an object of the higher-order formalism called a \textit{quantum tester}.

While quantum maps describes transformations of quantum states, higher-order operations (also called supermaps) describe transformations of quantum maps themselves. The equivalent of a POVM in this formalism is a quantum tester ---the most general higher-order transformation that maps quantum channels to a probability distribution, effectively ``measuring'' a quantum channel and yielding a classical outcome with some probability. As illustrated in Fig.~\ref{fig:setup}, a tester $T$ is equivalent to the concatenation of a state $\rho$ and a measurement $M$. Nevertheless, as we now explain, the Born rule in Eq.~\eqref{eq::prob_map} becomes linear in the tester variable, which is characterized by simple SDP constraints. We then exploit these two properties to efficiently optimize over the quantum strategies.

In order to express an estimation problem in terms of testers, we start by restating the problem using the Choi-Jamio\l{}kowski isomorphism~\cite{jamiolkowski72,choi75}. In this representation, a map $\Ecal_{\theta}: \Lcal(\Hcal^I) \to \Lcal(\Hcal^O)$ can be equivalently expressed as an operator $C_{\theta}\in\Lcal(\Hcal^I\otimes\Hcal^O)$, given by
$C_{\theta} = \sum_{ij} \ketbra{i}{j}\otimes\Ecal_{\theta}[\ketbra{i}{j}]$, called the Choi operator.

Using the Choi operator, the output state of the probe can be expressed as
\begin{equation}
    \rho_{\theta}=(\Ecal_{\theta}\otimes\text{id})[\rho] = \tr_I\Big( (C_{\theta}\otimes\id^\text{aux})\,(\rho^{\text{T}_I}\otimes\id^{O}) \Big), 
\end{equation}
where ${(\cdot)}^{\text{T}_I}$ denotes the partial transposition over the input space $\Hcal^I$. Then, the probability of obtaining outcome $i$, as in Eq.~\eqref{eq::prob_map}, can be equivalently written as
\begin{align}
    p(i|\theta) &= \tr\left[ \tr_I\Big((C_{\theta}\otimes\id^\text{aux})\,(\rho^{\text{T}_I}\otimes\id^{O})\Big) M_i\right]\\
    &= \tr\left[
    C_{\theta}\,\tr_\text{aux}\Big((\rho^{\text{T}_I}\otimes\id^{O})(\id^{I}\otimes M_i)\Big)\right].\label{eq::prob_choi}
\end{align}

We can now group the objects that constitute the quantum strategy, that is the state and the measurement, into a single object called the quantum tester~\cite{chiribella07,chiribella08memory,ziman08}. A tester $T=\{T_i\}_{i=1}^{N_O}$, $T_i\in\Lcal(\Hcal^I\otimes\Hcal^O)$ is a set of $N_O$ (standing for the ``number of outcomes'') operators defined as
\begin{equation}\label{eq::quantumtester}
    T_i \coloneqq \tr_\text{aux}\Big((\rho^{\text{T}_I}\otimes\id^{O})(\id^{I}\otimes M_i)\Big),
\end{equation}
which allows one to rewrite the probability of obtaining outcome $i$, in Eq.~\eqref{eq::prob_choi}, as simply
\begin{equation}\label{eq::prob_tester}
    p(i|\theta) = \tr\left(
    C_{\theta}\,T_i\right).
\end{equation}

The usefulness of this representation comes from the fact that, as shown in Ref.~\cite{chiribella07,chiribella08memory,ziman08}, testers have a simple mathematical characterization. More specifically, they obey the following set of necessary and sufficient conditions:
\begin{align}
    T_i &\geq 0 \ \ \ \forall \, i \label{eq::tester1} \\
    \sum_i T_i &= \sigma \otimes \id^O, \label{eq::tester2}
\end{align}
where $\sigma\in\Lcal(\Hcal^I)$, $\sigma\geq0$ and $\tr(\sigma)=1$. 
It is straightforward to see that every set of operators $T$ that satisfy Eq.~\eqref{eq::quantumtester} also satisfy Eqs.~\eqref{eq::tester1} and~\eqref{eq::tester2}. The converse is also true. Given any set of operators $T$ that satisfy Eqs.~\eqref{eq::tester1} and~\eqref{eq::tester2}, one can define a state $\rho$ and measurement $M=\{M_i\}_{i=1}^{N_O}$ according to
\begin{align} 
    \rho &\coloneqq\left(\id^{I}\otimes\sqrt{\sigma}\right)\sum_{ij}\ketbra{ii}{jj}\left(\id^{I}\otimes\sqrt{\sigma}\right)^\dagger \label{eq::realization_rho} \\
    M_i &\coloneqq  \left({\sqrt{\sigma}^{-1}}\otimes \id^{O}\right) \,T_i\, {\left({\sqrt{\sigma}^{-1}} \otimes \id^{O}\right)}^\dagger \ \ \ \forall \, i, \label{eq::realization_M} 
\end{align}
such that 
\begin{equation}
    \tr_\text{aux}\Big((\rho^{\text{T}_I}\otimes\id^{O})(\id^{I}\otimes M_i)\Big) = T_i  \ \ \ \forall \, i.
\end{equation}
The state $\rho$ and measurement $M$ are called a \textit{quantum realization} of the tester $T$. This realization is not unique, as different sets of states and measurements can lead to the same tester. However, crucially, different states and measurements that lead to the same tester will also yield the same probability distribution $\{p(i|\theta)\}_i$ in Eq.~\eqref{eq::prob_map}, and have the same performance in an estimation task.

Hence, the optimization of any linear function of $p(i|\theta)$ in Eq.~\eqref{eq::prob_tester} over a tester $T=\{T_i\}$ that satisfies Eqs.~\eqref{eq::tester1} and~\eqref{eq::tester2} is a semidefinite program, and its optimal tester is guaranteed to have a quantum realization in terms of a quantum state and measurement. Importantly, once the optimal quantum strategy (i.e. tester) is found, the corresponding optimal state and measurement can be easily determined using Eqs.~\eqref{eq::realization_rho} and~\eqref{eq::realization_M}. 

Notice that, while a tester is a set of operators that act only on the input and output space of the channel $C_{\theta}$, its quantum realization may require an auxiliary system. This implies that the optimal quantum strategy may require entanglement between the target and auxiliary systems, and a global measurement that acts on both of these systems. The dimension of the auxiliary space is bounded to be at most the dimension of $\Hcal^I$, as established by the explicit construction of $\rho$ in Eq.~\eqref{eq::realization_rho}. The auxiliary system can also be interpreted as a (quantum) memory. Hence, by optimizing over testers, one is effectively optimizing over all possible quantum strategies, including those that may require memory/entanglement for their implementation. 

However, certain experimental limitations might induce a situation in which it is necessary to design a quantum strategy that does not require entanglement for its implementation, or a means to certify whether entanglement is indeed advantageous in a given estimation task. In App.~\ref{app::noentanglement} we provide details on how quantum strategies that do not require entanglement can be approximated with SDPs. Moreover, in Sec.~\ref{sec::example_thermometry} we provide an example of a temperature estimation problem in which our methods demonstrate a clear gap between the performance of strategies operating with and without entanglement.

\section{Optimal tester for metrology via semidefinite programming}

Using quantum testers, we can now rewrite the score of an estimation problem in Eq.~\eqref{eq::continuous_score} as 
\begin{equation}
    \cS = \sum_{i=1}^{N_O}  \int d\theta \, p(\theta) \, r(\theta, \hat{\theta}_i) \, \tr(C_{\theta}\,T_i).
\end{equation}
Now, to find the optimal score of a given estimation task, the optimization of $\cS$ over the triplet $\{\rho,~\{M_i\}_i,~\{\hat{\theta}_i\}_i\}$ can be substituted for an optimization over the pair $\{\{T_i\}_i,~\{\hat{\theta}_i\}_i\}$. 

We may express all dependencies of the score ${\cS}$ on the estimates $\{\hat{\theta}_i\}_{i=1}^{N_O}$ with a set of operators $\{X(\hat{\theta}_i)\}_{i=1}^{N_O}$, $X(\hat{\theta}_i)\in\Lcal(\Hcal^I\otimes\Hcal^O)$, which are given by an integral over the parameter $\theta$, defined as 
\begin{equation}\label{eq::Xintegral}
    X(\hat{\theta}_i) \coloneqq \int d\theta \, p(\theta) \, r(\theta, \hat{\theta}_i) \, C_{\theta} \ \ \ \forall \, i.
\end{equation}
These operators encompass all the given information about the task (prior distribution, cost function, and channels in which the parameter is encoded) that does not depend on the quantum strategy. Expressed in terms of these operators, the score is simply
\begin{equation}\label{eq:exact_score_pair}
    \cS = \sum_{i=1}^{N_O} \tr \left(X(\hat{\theta}_i)\,T_i\right).
\end{equation}

For any given set of fixed estimates $\{\hat{\theta}_i\}$, the optimization of the score is given by either a maximization or minimization (depending on the character of the cost function) of $\cS$ over all testers $T$. Taking maximization for instance, we have that
\begin{equation}\label{eq::optS}
   \max_{\{T_i\}} \cS = \max_{\{T_i\}} \sum_{i=1}^{N_O} \tr \left(X(\hat{\theta}_i)\,T_i\right),
\end{equation}
is the optimal score. The optimization over testers includes the constraints of Eqs.~\eqref{eq::tester1} and~\eqref{eq::tester2}. Since testers $T=\{T_i\}$ are sets of positive semidefinite operators characterized by linear constraints, the above optimal score can be efficiently computed using SDP. Once again, the optimal tester is guaranteed to have a quantum realization, hence for any optimal solution of $T$ that the SDP should return, there exist a probe state $\rho$ and measurement $M$ that can realize it; they constitute the optimal quantum strategy for the given estimators. 

Notice that this can be straightforwardly generalized to the multiparameter regime as well. In App.~\ref{app::multiparameter} we provide more details on this case, while in Sec.~\ref{sec::example_su2} we present an example of the application of our methods to the multiparameter problem of SU(2) estimation.

It is now clear that given the knowledge of the estimator values $\hat \theta_i$ and the operators $X(\hat \theta_i)$ one can find the optimal tester $T$ efficiently. The remaining difficulties thus are:
\begin{enumerate}[label=(\arabic*)]
    \item Finding the optimal estimators $\{\hat{\theta}^*_i\}$ leading to the optimal score
    \begin{align}\label{eq:opt_score}  \cS^* \coloneqq  \underset{\{T_i\}, \{{\hat \theta}_i\}}{\max} 
    ~\cS, \qquad
    \{\{T_i^*\}, \{\hat{ \theta}_i^*\}\} \coloneqq 
    \underset{\{T_i\}, \{{\hat \theta}_i\}}{\arg\max} 
    ~\cS.
    \end{align}
     \item Computing the integral in Eq.~\eqref{eq::Xintegral}.
\end{enumerate}
In the following, we construct three different approaches to tackle both of these problems.

\section{Parameter discretization and estimator optimization }\label{sec::methods}

In situations where the optimal estimators are unknown, or the integral in Eq.~\eqref{eq::Xintegral} cannot be calculated exactly, an approximation of the optimal score in Eq.~\eqref{eq::optS} can still be computed with SDP. This can be achieved by first discretizing the parameter $\theta$ to a finite number of hypotheses, thereby mapping the original parameter estimation task onto one closely resembling channel discrimination. 

Concretely, let us choose a discretization of $\theta$ such that $\theta \mapsto \{\theta_k\}_{k=1}^{N_H}$, where $N_H$ (standing for the ``number of hyphotheses'') is the total number of different values assigned to $\theta$. We can then define a prior distribution over the new hypotheses as
\begin{equation}
    p(\theta_k) \coloneqq \frac{p(\theta = \theta_k)}{\sum_{k=1}^{N_H} p(\theta = \theta_k)},
\end{equation}
which is computationally straightforward and has the advantage of giving a valid probability distribution.

Now, let's define the discrete equivalent of the operators in Eq.~\eqref{eq::Xintegral} as $\{\widetilde{X}(\hat{\theta}_i)\}_{i=1}^{N_O}$, where
\begin{equation}\label{eq::Xsum}
    \widetilde{X}(\hat{\theta}_i) \coloneqq \sum_{k=1}^{N_H} p(\theta_k) \, r(\theta_k, \hat{\theta}_i) \, C_{\theta_k} \ \ \ \forall \, i.
\end{equation}

Hence, the approximate score $\widetilde{\cS}$ can be expressed as 
\begin{equation}\label{eq::discreteS}
    \widetilde{\cS} \coloneqq \sum_{i=1}^{N_O} \tr \left(\widetilde{X}(\hat{\theta}_i)\,T_i\right).
\end{equation}
The value of $\widetilde{\cS}$ will depend on a chosen discretization $\{\theta_k\}$ of the continuous parameter $\theta$---the finer the discretization, the better the approximation. Hence, for a given discretization $\{\theta_k\}$, the optimum score is given by either maximizing or minimizing $\widetilde{\cS}$ again over the pair $\{\{\hat{\theta}_i\}_i,\{T_i\}_i\}$ of estimates and testers. 

In what follows we propose three different methods, all based on semidefinite programming, with which this approximation can be computed. 
 
\subsection{Method 1: Approximating metrology with channel discrimination}\label{sec::method1}

The first approach we propose is heavily based on the problem of channel discrimination~\cite{kitaev97}. Its starting point is the realization that, without loss of generality, we may restrict ourselves to testers with as many measurement outcomes as there are hypotheses to be distinguished. In the context of our discretized parameter estimation problem, this amounts to setting $N_O=N_H$; essentially, there is no advantage in increasing the number of measurement outcomes beyond the number of different values in the discretization of $\theta$.  The second simplification is to choose the values of the estimates $\{\hat{\theta}_i\}$ to be the same as the values in the discretization $\{\theta_k$\}, in such a way that each measurement outcome $i$ is directly associated to a value $\hat{\theta}_i=\theta_i$. Choosing the values of the potential estimates of the parameter to correspond to the values in the discretization of the parameter reduces the estimation problem to a discrimination problem. In this case, the task can be interpreted as determining the ``classical'' label $k$ that is encoded via the values of $\theta_k$ in the channel $C_{\theta_k}$. In this case, the set of operators $\{\widetilde{X}(\hat{\theta}_i)\}$ becomes 
\begin{equation} \label{eq:m1_x}
    \widetilde{X}(\hat{\theta}_i) = \widetilde{X}(\theta_i) = \sum_{k=1}^{N} p(\theta_k) \, r(\theta_k, \theta_i) \, C_{\theta_k} \ \ \ \forall \, i,
\end{equation}
where $N=N_O=N_H$, and the approximate score $\widetilde{\cS}$ becomes
\begin{equation} \label{eq:score_m1}
    \widetilde{\cS} = \sum_{i=1}^{N} \tr \Big(\widetilde{X}(\theta_i)\,T_i\Big).
\end{equation}
For this fixed values of the discretization $\{\theta_i\}_{i=1}^N$, the optimum value of $\widetilde{\cS}$ over all testers $T$ is an SDP.

This approach circumvents problem (1), of finding the optimal estimators, by setting them to be the same values used in the discretization of the continuous parameter $\theta$; and problem (2), of computing the integral in Eq.~\eqref{eq::Xintegral}, by discretizing it. In principle, the higher the number of values in the discretization of $\theta$, the closer the estimates are to the optimal estimator. The advantage then is that the optimal score can be found with a single SDP that needs to optimize only over the quantum strategies. The drawback, on the other hand, is that to achieve a good approximation of the optimal estimator, a high number $N$ of values in the discretization are necessary, and since this number is directly associated to the number of measurement outcomes in the quantum strategy, the problem can eventually become intractable numerically and experimentally. In practice, however, as demonstrated in our examples in Sec.~\ref{sec::examples}, this method yields very good results with a value of $N$ that can still be straightforwardly handled numerically. 

Nevertheless, our next approach is designed to overcome this problem as well.

\subsection{Method 2: Parameter discretization with optimal estimator}\label{sec::method2}
One possible way to overcome computational challenges is to fix the number of measurement outcomes $N_O$, and hence the number of tester elements, to a value that is computationally (and experimentally) tractable and increase the value of $N_H$ far beyond that. Since, in this case, the complexity of the problem does not depend on $N_H$, the discretization of $\theta$ can be arbitrarily fine. However, because the number of values in the discretization of $\theta$ can far surpass the number of measurement outcomes, the association of one estimate $\hat{\theta}_i$ to each discretization value $\theta_k$ is no longer possible. Therefore, the problem of choosing a ``good'' set of estimates $\{\hat{\theta}_i\}$ is crucial. 

Let us start by assuming that the optimal estimator
is known to be $\{\hat{\theta}^*_i\}$. Then, the operators $\{\widetilde{X}(\hat{\theta}^*_i)\}$ amount to
\begin{equation}
    \widetilde{X}(\hat{\theta}^*_i) \coloneqq \sum_{k=1}^{N_H} p(\theta_k) \, r(\theta_k, \hat{\theta}^*_i) \, C_{\theta_k}, \ \ \ \forall \, i,
\end{equation}
for a fixed discretization $\{\theta_k\}_{k=1}^{N_H}$, which can now in principle contain an arbitrarily high number of values $N_H\gg N_O$. 
The approximate score $\widetilde{\cS}$ then becomes
\begin{equation}\label{eq:tilde_S_II}
    \widetilde{\cS} = \sum_{i=1}^{N_O} \tr \Big(\widetilde{X}(\hat{\theta}^*_i)\,T_i\Big).
\end{equation}
The optimization of $\widetilde{\cS}$ over the quantum strategy is then given by an SDP.

This approach essentially takes care of problem (2), of numerically computing the integral in Eq.~\eqref{eq::Xintegral}, by discretizing the parameter $\theta$ in an arbitrarily fine manner, while maintaining the number of measurements low enough to decrease the computational demand of the SDP. Hence, it is better suited for a situation in which the optimal estimator is known. It can nevertheless also be applied to a problem in which only a good guess for the optimal estimator is known, in which case the solution will be an approximation of the optimal score. Otherwise, to overcome problem (1) of finding the optimal estimator in the first place, we combine this approach with an estimator optimization in a seesaw algorithm, detailed in the following.

\subsection{Method 3: Parameter discretization with estimator optimization}\label{sec::method3}

This final approach consists of a seesaw between two optimization problems---which are not necessarily SDPs---that will approximate an optimization over both the quantum strategy and the values of the estimates.

A seesaw is an iterative method that alternates between two optimization problems, using the solution of one as the input of the other. In our case, the first optimization problem is the SDP of the previous approach (Method 2). Namely, given $\{\hat{\theta}_i\}_{i=1}^{N_O}$, 
\begin{equation}\label{eq::seesaw1}
    \max_{\{T_i\}}\ \ \sum_{i=1}^{N_O} \tr \Big(\widetilde{X}(\hat{\theta}_i)\,T_i\Big), 
\end{equation}
where $\{T_i\}$ is an $N_O$-outcome tester.

The second optimization problem will then be one that, for a fixed tester $\{T_i\}$, taken to be the optimal tester of the previous SDP, optimizes over the values $\{\hat{\theta}_i\}$ of the estimates. Namely, given $\{T_i\}_{i=1}^{N_O}$,
\begin{equation}\label{eq::seesaw2}
    \max_{\{\hat{\theta}_i\}}\ \ \sum_{i=1}^{N_O} \tr \Big(\widetilde{X}(\hat{\theta}_i)\,T_i\Big),  
\end{equation}
where $\{\hat{\theta}_i\}$ are $N_O$ possible values of $\theta$.

Whether the problem in Eq.~\eqref{eq::seesaw2} is an SDP will depend on whether the reward function $r(\theta_k, \hat{\theta}_i)$ is linear on $\{\hat{\theta}_i\}$. In practice, this will often not be the case. Nevertheless, in some cases this problem can be solved analytically---depending on the form of the reward function, the optimal estimator may be known or it may be found by standard Lagrangian optimization methods. In other cases, heuristic optimization methods may be applied.

This iterative method, although even for a fixed discretization $\{\theta_k\}$ does not necessarily converge to the optimal value of $\widetilde{\cS}$, in practice leads to very good approximations. A relevant point here is that, assuming a situation where the seesaw does converge to the optimal estimator, one may restrict themselves without loss of generality to a maximum number of outcomes $N_O$ that is related to the extremality properties of the tester. In principle, since (i) the set of testers $T=\{T_i\}$ is convex and (ii) the function $\widetilde{\cS}$ is linear on each tester element $T_i$, the maximum (or minimum) of $\widetilde{\cS}$ will be achieved by an extremal tester. Analogously to extremal POVMs~\cite{dariano05}, extremal testers have at most $d^2$ (non-zero) elements, where $d$ is the dimension of the space upon which the tester (or POVM) elements act. Hence, the number of outcomes $N_O$ in the seesaw can be fixed to be at most $N_O\leq(d_I\times d_O)^2$, where $d_{I(O)}\coloneqq\text{dim}(\Hcal^{I(O)})$, since, for optimal estimators, there is no advantage in optimizing over non-extremal testers. This fact also holds for Method 2 if one is guaranteed to know the optimal estimator. 

Furthermore, if the cost function is the mean squares error $r_\text{MSE}$, then the optimal measurement will be projective (see Appendix A in Ref.~\cite{macieszczak2014bayesian}), and hence the optimal tester will have at most $(d_I\times d_O)$ outcomes. We present a case study in Sec.~\ref{sec::example_thermometry}, which concerns the problem of thermometry, that precisely falls in this case.

\subsection{Convergence of the Methods}

In all three methods above, we encounter some error due to discretization of the integral in finite hypotheses, as well as sub-optimality due to our choice of estimators. The discretization error is expected to vanish as $N_H$ increases, since all three methods are based on approximating an integral with a Riemannian sum with an error that vanishes as $1/N_H$. 
As for the sub-optimality, let us define the best approximate score ${\widetilde{\cS}}^*$ similarly to Eq.~\eqref{eq:opt_score} but for the approximate score defined in Eq.~\eqref{eq::discreteS}. For large enough $N_H$, and when the cost function is supposed to be maximised, it means that ${\widetilde{\cS}}^* \leq \cS^{*}$ while for cost functions that are supposed to be minimised, it means that ${\widetilde{\cS}}^* \geq \cS^{*}$.
The sub-optimality roots from the fact that, none of the methods simultaneously optimise over both $\{\hat \theta_i\}$ and $\{T_i\}$. Each of the three methods, however, deals with sub-optimality differently. In all three methods one has 
\begin{equation}
    |\widetilde{\cS}^* - \cS^*| = {\cal O}\left(\frac{1}{N_O}\right)
\end{equation}
and thus one can guarantee convergence by choosing $N_O\gg 1$. In Appendix~\ref{app:convergence} we rigorously derive the convergence for arbitrary cost functions and furthermore show that for certain cost functions that we will later use in the case studies (Examples 1. and 2.) the convergence is even faster, i.e., $|\widetilde{\cS}^* - \cS^*| = {\cal O}(1/N_O^2)$. When we cannot arbitrarily increase $N_O$, Methods 2 and 3 come to the rescue. In particular, if a priori we know what the optimal estimators are, then Method 2 allows to find the optimal testers in one shot. However, it is rarely the case that we do know the optimal estimators a priori. Nonetheless, as we see in the examples below, Method 2 typically finds sub-optimal solutions that are very close to the optimum. Method 3, on the other hand, adds a powerful layer of optimization based on a seesaw between the estimators and testers, and therefore has a higher chance of finding the optimal protocol even with $N_O = (d_I\times d_O)^2$.

\section{Case studies}\label{sec::examples}

Our methodology for solving the Bayesian parameter estimation problem using higher-order operations offers numerous advantages over conventional techniques in quantum metrology. 
By proposing to optimize over the input state and measurement with a single SDP, and combining this with effective heuristics for the joint optimization of the quantum strategy and estimator, we overcome the longstanding challenges of the Bayesian approach. Our approach provides a comprehensive and versatile set of techniques that can be applied to any Bayesian estimation problem, setting it apart from most existing methods in the literature.

The key strength of our approach lies in its ability to handle a wide range of estimation problems, without being limited to specific error quantifiers. This universality allows our method to be seamlessly applied to any estimation scenario. Moreover, the techniques we described here are equally effective for single parameter and multiparameter estimation tasks. Finally, unlike most techniques in the Bayesian approach, our methods are not bound by the type of dynamics used to encode the parameter. Whether the parameter is encoded via a unitary evolution (e.g. phase estimation), or a more complex open system dynamics resulting from the probe's interaction with a thermal environment (e.g. quantum thermometry), our approach can be systematically applied and, as we show in the following, delivers consistent results which are very close to the optimal values.

We now delve into the practical application of our methods and explore how they can be applied to determine the optimal estimation strategy for various scenarios encountered in quantum metrology. The examples are deliberately chosen to be cover a wide range of different problems to demonstrate the versatility of our methods.

\subsection{Example 1: Paradigmatic example -- Local phase estimation}\label{sec::example_qutritphase}

We start with a paradigmatic task in quantum metrology, namely the single-parameter unitary phase estimation. In this problem a single parameter $\theta \in [0,2\pi)$ is encoded in an $n$-qubit quantum system via a local unitary channel $\mathcal{E}_{\theta}[\cdot] = U_{\theta} (\cdot) U_{\theta}^\dag$ with
\begin{equation}\label{eq:phase1}
    U_{\theta} = e^{-i \theta S_z},
\end{equation}
where $S_z$ is a collective spin operator $S_z = \frac{1}{2} \sum_{i=1}^n \sigma_z^{(i)}$ and $\sigma_z^{(i)}$ is the Pauli-Z matrix of the $i$-th qubit. Due to the intrinsic symmetry of the problem, every state of the $n$-qubit system can be effectively described using an $(n+1)-$dimensional Hilbert space, i.e. the symmetric subspace~\cite{harrow2013church}. Consequently, the $n$-qubit phase estimation problem can be equivalently mapped into a phase estimation problem of a $d-$dimensional system with $d = n + 1$. In this representation the generator of the dynamics, $S_z$, expressed in the computational basis $\{\ket{0},\ket{1},\dots, \ket{n}\}$ is given by $S_z = \sum_{i=0}^{n} i \ketbra{i}$ \cite{knysh2014true}. 

For this example, we take a typical reward function in phase estimation, which takes into account the cyclicity of its parameter space, given by
\begin{align}\label{eq::cost_cos2}
    r(\theta, \hat{\theta}_i) = \cos^2\left(\frac{\theta - \hat{\theta}_i}{2}\right).
\end{align}
We also choose two different priors, one given by a uniform distribution according to 
\begin{equation}\label{eq::uniformprior}
    p(\theta) = \frac{1}{\theta_{\max}-\theta_{\min}},
\end{equation}
where $\theta_{\text{min}}=0$ and $\theta_{\text{max}}=2\pi$ are respectively the minimal and maximal values of the parameter. The other distribution is given by a Gaussian distribution, according to 
\begin{equation}\label{eq::gaussianprior}
    p(\theta) = \frac{1}{{\cal N}} \exp{\frac{(\theta - \mu)^2}{(2\sigma^2)}},
\end{equation}
where ${\cal N}$ is the normalization factor. We set the mean $\mu=\pi$ and the deviation $\sigma=1$. For the discretization of the parameter $\theta$ and initial value of the estimators, we fix
\begin{equation}\label{eq::discretization}
    \theta_k  = \frac{\theta_{\text{max}} - \theta_{\text{min}}}{N_H} (k-1) \quad \forall \, \, k \in \{1, \ldots, N_H\},
 \end{equation}
and \begin{equation}\label{eq::estimators}
    \hat\theta_i  = \frac{\theta_{\text{max}} - \theta_{\text{min}}}{N_O} (i-1) \quad \forall \, \, i \in \{1, \ldots, N_O\},
\end{equation}
respectively.

\begin{figure}
\begin{center}
	\includegraphics[width=\columnwidth]{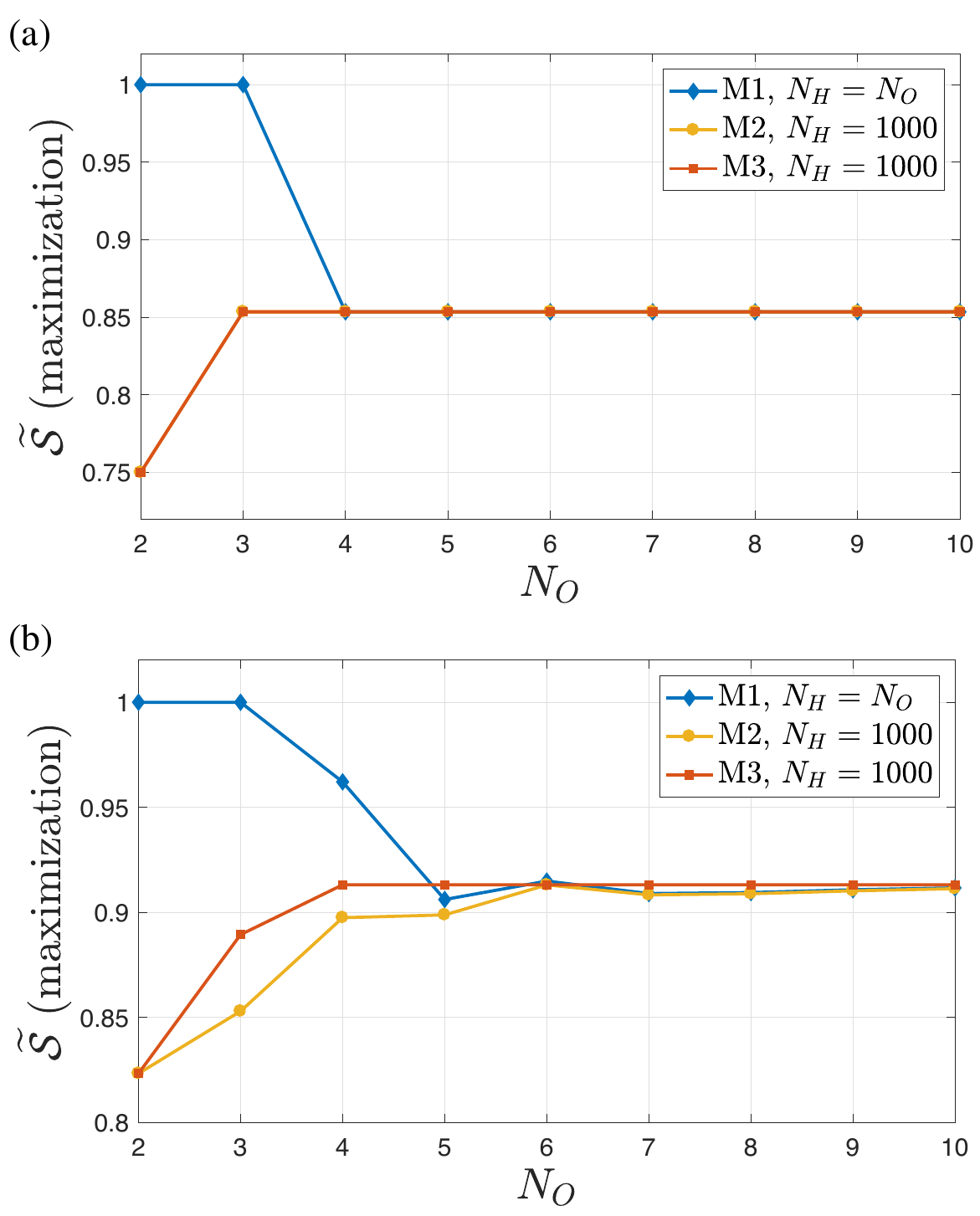}
    \caption{\textbf{Local phase estimation (Example 1).} The maximum approximate score $\widetilde{\mathcal{S}}$ in a local $n=2$ qubit phase estimation problem.  Each panel shows the scores corresponding to Methods M1, M2, and M3 as a function of the number of outcomes $N_O\in\{2,\ldots,10\}$ for different prior distributions of the local phase: panel $(a)$ corresponds to the case of uniform prior, while $(b)$ corresponds to a Gaussian prior. The phase parameter ranges from $\theta_\text{min}=0$ to $\theta_\text{max}=2\pi$. The considered cost function is the cosine squared in Eq.~\eqref{eq::cost_cos2}.}
\label{fig::qutritphase}
\end{center}
\end{figure}

We now discuss how to apply each of our methods to infer the optimal protocol in this case and present the results obtained for the problem of $(n=2)$-qubit phase estimation, plotted in Fig.~\ref{fig::qutritphase}.

\textbf{{Method 1.}} To apply Method 1, we simply set $N_H=N_O=N$, implying $\hat{\theta}_i=\theta_i$. For the results plotted in Fig.~\ref{fig::qutritphase}, we take $N\in\{2,\ldots,10\}$.

\textbf{Method 2.} Here we fix the number of outcomes $N_O\in\{2,\ldots,10\}$ and set the number of hypotheses to be $N_H = 1000 \gg N_O$. We then discretize the parameter and set the estimators according to Eqs.~\eqref{eq::discretization} and~\eqref{eq::estimators}. For the case of a uniform prior (Eq.~\eqref{eq::uniformprior}, Fig.~\ref{fig::qutritphase}(a)), these are expected to be the optimal estimators. For the case of a Gaussian prior (Eq.~\eqref{eq::gaussianprior}, Fig.~\ref{fig::qutritphase}(b), these estimators are not expected to be optimal, but are nonetheless used in Method 2, serving as a starting point for the estimator optimization in Method 3. 

\textbf{Method 3.} For this method, we take the solution for the testers found using Method 2 for each $N_O$ as a starting point, and then optimize over the estimator. We prove in App.~\ref{app:convergence} that in this case the optimal estimators are given as a function of the optimal tester, according to 
\begin{align}\label{eq:optimal_est_cos2}
    \hat \theta_i^* = \begin{cases}
    \arctan\left(
    \frac{\avg{\sin(\theta_k)}^{(i)}}{\avg{\cos(\theta_k)}^{(i)}}\right) & \avg{\cos(\theta_k)}^{(i)}\geq 0 \\
    \arctan\left(
    \frac{\avg{\sin(\theta_k)}^{(i)}}{\avg{\cos(\theta_k)}^{(i)}}\right) +\pi & \avg{\cos(\theta_k)}^{(i)}< 0 
\end{cases},
\end{align}
where 
\begin{equation}\label{eq::avgf}
    \avg{f(\theta_k)}^{(i)} = \sum_k p(\theta_k|i)f(\theta_k) = \sum_k\frac{p(\theta_k)\tr(C_{\theta_k} T_i)}{p(i)} f(\theta_k) 
\end{equation} 
with $p(i)=\sum_k p(\theta_k)\tr(C_{\theta_k} T_i)$. Note that with the definition of Eq.~\eqref{eq:optimal_est_cos2} the range of the estimator is $[-\frac{\pi}{2}, \frac{3\pi}{2}]$, instead of $[0,2 \pi]$ as defined initially, this has no effect on the expected reward and can be resolved by adding $2\pi$ if $\hat \theta_i^* <0$. Hence, the second step in the see-saw is solved analytically, and in each round of the seesaw we update the value of the estimators according to the expression above, as a function of the testers found by the SDP in the first step. Here and in the following examples, we iterate these two steps until the gap between the value of the score in subsequent rounds is smaller than $10^{-6}$.

\textbf{Results.} In Fig.~\ref{fig::qutritphase} we plot the maximal approximate scores $\widetilde{S}$ obtained via the three methods outlined above. In the case of the uniform prior (Fig.~\ref{fig::qutritphase}(a)) we observe that the approximate score very quickly reaches the optimal one, i.e. $\widetilde{\mathcal{S}} \approx \mathcal{S} = \frac{1}{2}(1+\cos(\pi/4))$ which was formerly obtained with alternative methods~\cite{berry2000optimal,bartlett2007reference}. All three methods converge very quickly to this solution, already for $N_O = 3$ outcomes, using Method 2 and 3, and for $N_O = 4$ outcomes, using Methods 1. Notice also that, since we start already at the optimal estimator, we observe that for all $N_O$ there is no advantage of applying the seesaw in Method 3, since Method 2 already returns the optimal solution. For the case of the Gaussian prior (Fig.~\ref{fig::qutritphase}(b)), we see that again Method 3 converges to a stable value of $\widetilde{\cS}$ with $N_O=4$ outcomes. Here we can see an initial difference between Methods 1 and 2, which take a fixed estimator, and Method 3, which optimizes over the estimator. Nevertheless, all methods quickly converge to approximately the same value, at $N_O=10$.

\subsection{Example 2: Non-unitary evolution -- Thermometry}\label{sec::example_thermometry}

Let us now discuss a different instance of single parameter estimation, namely thermometry~\cite{Mehboudi_2019, DePasquale2018}. In this case, the unknown parameter is the temperature $\theta$ of a sample (or a thermal bath) that is resting at thermal equilibrium, and it is encoded in the probe using a non-unitary quantum channel. 
We consider the probe to be a two level system (qubit) which is potentially entangled with an auxiliary system---that does not undergo the non-unitary dynamics. At the initial time $t = 0$, the probe and the sample which are initially uncorrelated start to interact. After some fixed time $t$, the probe and the auxiliary system will be jointly measured to infer the temperature of the bath. The probe's reduced state $\rho^p_{\theta} = {\rm tr}_A[\rho_{\theta}]$ evolves according to a standard Markovian quantum master equation~\cite{kossakowski1972quantum,gorini1976completely,lindblad1976generators,breuer2002theory}, i.e. 
\begin{align}\label{eq:master_equation}
    {\dot \rho}^p_{\theta}(t) & = -i[H,\rho^p_{\theta}(t)] + \Gamma_{\rm in} {\cal D}[\sigma_+]\rho^p_{\theta}(t) +\Gamma_{\rm out} {\cal D}[\sigma_-]\rho^p_{\theta}(t),
\end{align}
where $H=\epsilon \ket{1}\bra{1}$ is the Hamiltonian of the probe, $\sigma_-= \ket{0}\bra{1}$ and $\sigma_+= \ket{1}\bra{0}$ are the jump operators, and ${\cal D}[A]\rho^p_{\theta} = A\rho^p_{\theta} A^{\dagger}-\sfrac{1}{2}\left\{ A^{\dagger} A, \rho^p_{\theta} \right\}$ is the dissipator superoperator which captures the effect of the environment on the probe. The dissipation rates $\Gamma_{\rm in}$ and $\Gamma_{\rm out}$ are the only temperature dependent parts of the dynamics and are responsible for encoding the parameter. For a bosonic/fermionic environment, we have $\Gamma_{\rm in} = J(\epsilon) N_{B/F}$ and $\Gamma_{\rm out} = J(\epsilon) (1 \pm N_{B/F})$---where minus sign should be used for fermions, and positive sign for bosons---with $J(\epsilon)$ being the bath spectral density while $N_{B/F}$ is the occupation number for the bosonic or fermionic bath, defined as $N_B = (e^{\epsilon/\theta}-1)^{-1}$ and $N_F = (e^{\epsilon/\theta}+1)^{-1}$, respectively. In what follows, we focus on the bosonic bath, however our methods can be applied to the fermionic case as well.

The evolution specified by Eq.~\eqref{eq:master_equation} generates an effective quantum channel $\mathcal{E}_{\theta}(t)$ that imprints the temperature into the probe's state (see App.~\ref{app:proof_master_eq} for the explicit expression). Note that in our notation we keep the time dependence because we are also interested in the optimal protocol at different times. 

As for the cost function, we use the MSE
\begin{equation}\label{eq::cost_mse}
    r(\theta, \hat{\theta}_i) = (\theta - \hat{\theta}_i)^2,
\end{equation}
while the prior distribution $p(\theta)$ is uniform and given by Eq.~\eqref{eq::uniformprior}, where we set $\theta_\text{min}=0.1$ and $\theta_\text{max}=2$ as the minimum and maximum values of the temperature.

We discretize the temperature parameter $\theta$ and fix the estimators according to Eqs.~\eqref{eq::discretization} and~\eqref{eq::estimators}, respectively. We evaluate the thermometry problem for 100 different time steps, evenly distributed between $t=0$ and $t=1$. 

Let us now discuss how to approach this problem using each of the three methods presented in this work. 

\begin{figure}
\begin{center}
	\includegraphics[width=\columnwidth]{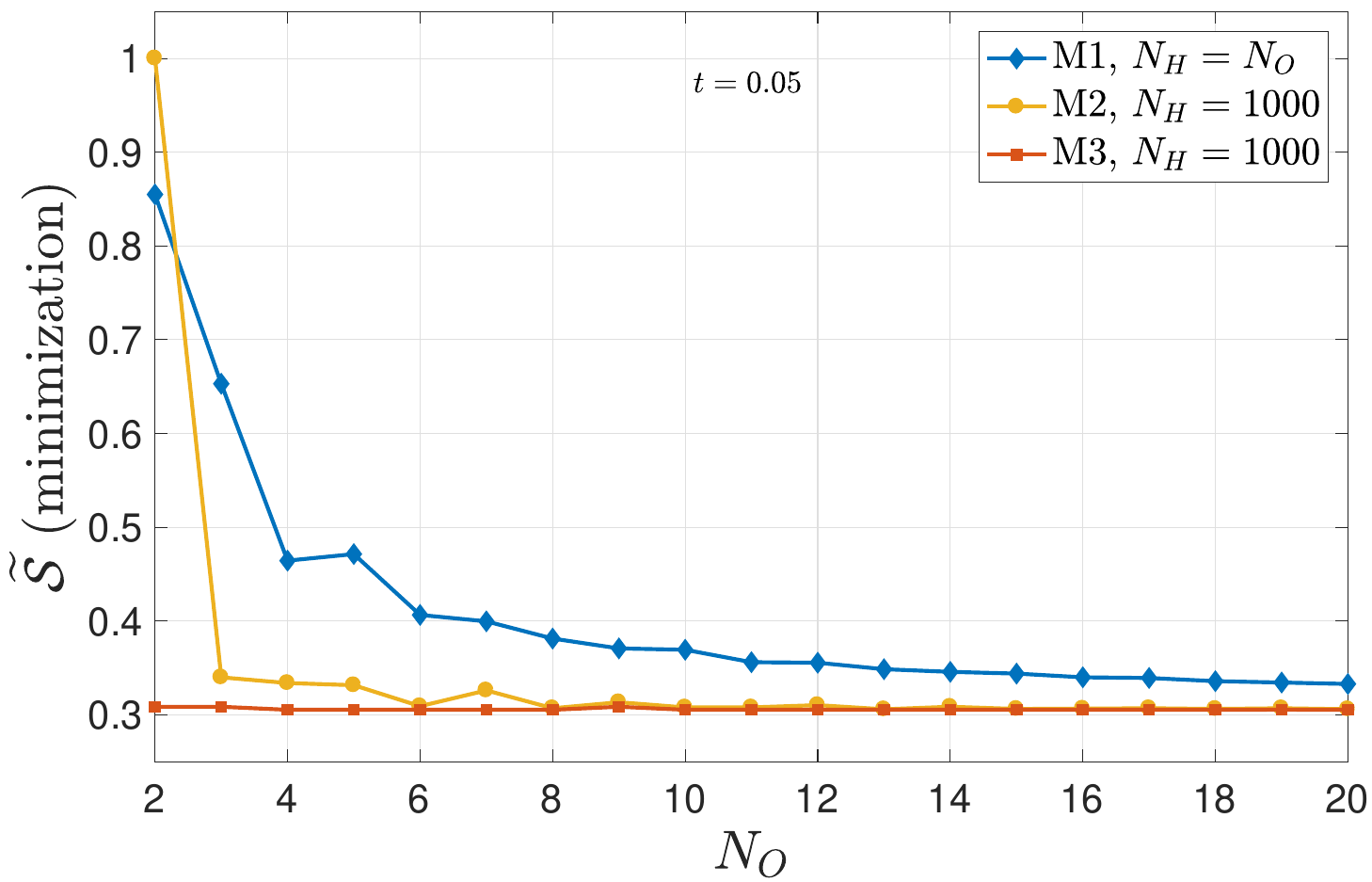}
    \caption{\textbf{Thermometry (Example 2).} The minimum approximate score $\widetilde{\mathcal{S}}$ in the finite-time temperature estimation problem $\widetilde{\mathcal{S}}$, renormalized by the maximum value of $\widetilde{\mathcal{S}}$ in the plot. The temperature $\theta$ is encoded via a qubit non-unitary evolution specified by Eq.~\eqref{eq:master_equation} acting for an amount of time $t < \infty$, here shown for a fixed time $t = 0.05$. The plot shows the different scores corresponding to Methods M1, M2, and M3 as a function of the number of outcomes $N_O\in\{2,\ldots,10\}$ for a uniformly distributed prior in a temperature parameter range of $\theta_{\text{min}} = 0.1$, $\theta_{\text{max}} = 2$. The considered cost function is the MSE in Eq.~\eqref{eq::cost_mse}. The remaining parameters chosen are $\epsilon = 0.1$ and $J(\epsilon) = 2$.}
    \label{fig::allmethods_allNo_fixedtime}
\end{center}
\end{figure}

\begin{figure}
\begin{center}
	\includegraphics[width=\columnwidth]{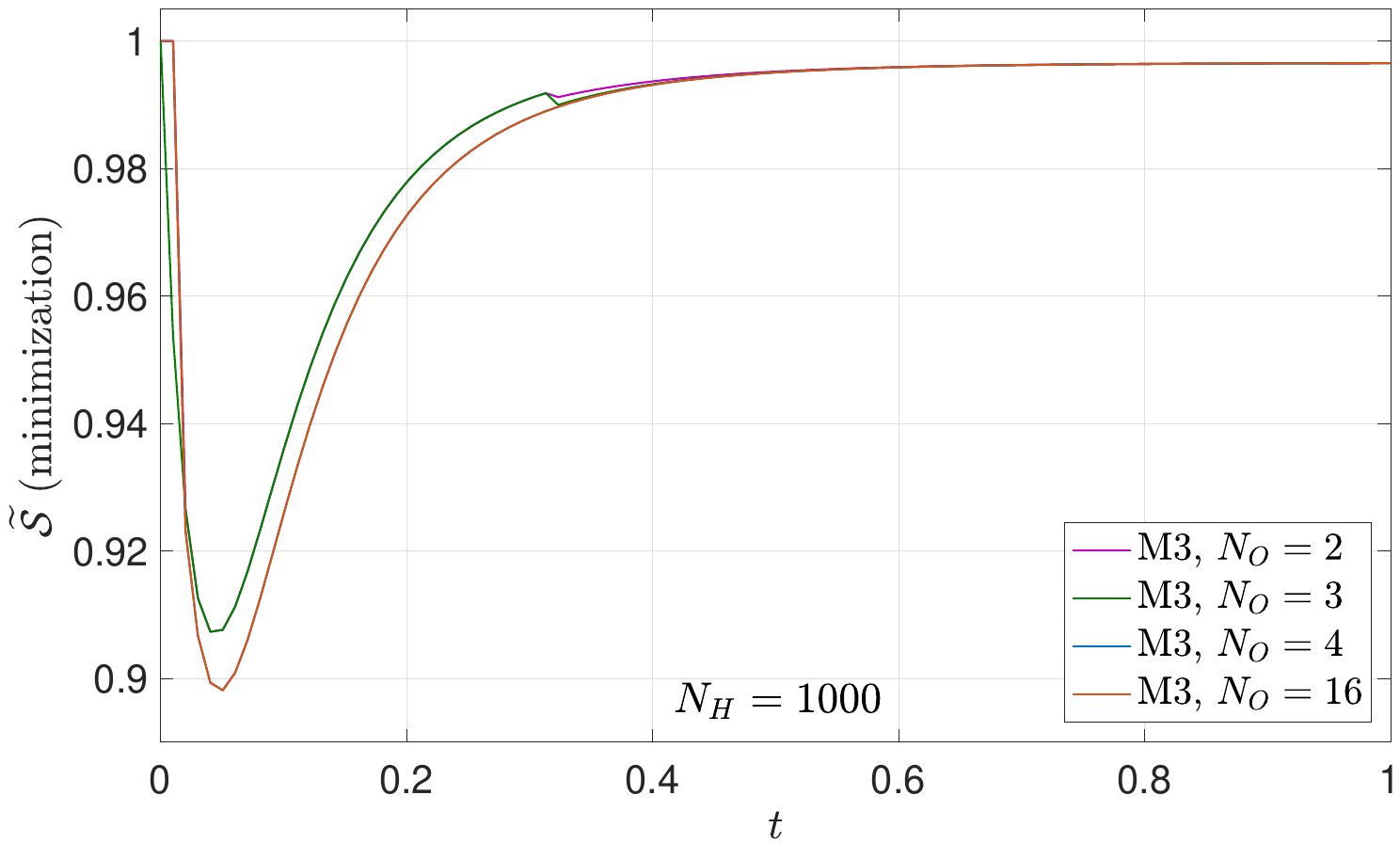}
	\caption{\textbf{Thermometry (Example 2): Optimality of projective measurements.} The approximate score $\widetilde{S}$ is computed using Method 3 as a function of time $t$ for   different values of $N_O$. The parameters are chosen as in Fig.~\ref{fig::allmethods_allNo_fixedtime}. The inset plot is a log-log plot of the same curves. All values are renormalized by the maximum value of $\widetilde{\mathcal{S}}$ in the plot. Since the cost function is the MSE, in this case projective measurements (which have at most $d_I\times d_O$ outcomes) are optimal. Indeed, we observe that increasing $N_O$ beyond $4$ does not change the value of the score.}
\label{fig::M3_differentNo_alltimes}
\end{center}
\end{figure}

\textbf{Method 1.} To apply Method 1, we again simply set $N_H=N_O=N$, implying $\hat{\theta}_i=\theta_i$. For the results presented here we take $N\in\{2,\ldots,20\}$.
 
\textbf{Method 2.} Here we fix the number of outcomes $N_O\in\{2,\ldots,20\}$ and set the number of hypotheses to be $N_H = 1000 \gg N_O$. These values for the estimator are not expected to be optimal but are nevertheless used in Method 2, serving as an starting point for the estimator optimization in Method 3.

\textbf{Method 3.}
To apply the seesaw in Method 3, we again begin with the solution provided by Method 2 for each $N_O$ and $t$ as a starting point. For the thermometry problem, as was the case for the phase estimation problem, we can analytically express the optimal estimator as a function of the quantum strategy (tester). Since the score is quantified using the MSE, the optimal estimator is simply the mean over the posterior 
\begin{align} \label{eq:ex2_est}
    \hat{\theta}_i = \avg{\theta_k}
\end{align}
where $\avg{\theta_k}$ is given by Eq.~\eqref{eq::avgf}. Once again, in this example the first step of the seesaw consists in an optimization over the testers while the second step consists in reassigning a value to the estimators as a function of the testers found in the previous step, according to the above expression. 

\textbf{Results.} In Fig.~\ref{fig::allmethods_allNo_fixedtime} we compare the performance of the three methods outlined above as a function of the number of outcomes $N_O$ for a fixed time $t=0.05$. We observe that Methods 1 and 2 start from a relatively large $\widetilde{S}$, which is now being minimized, that gradually decreases with increasing $N_O$, while Method 3 already starts at a value of $\widetilde{S}$ close to where it will converge. While Method $2$ quickly converges to the same values of $\widetilde{S}$ of Method 3 with increasing $N_O$, at $N_O = 20$ the approximate score predicted by Method $1$ is still somewhat above the corresponding one obtained using Method $3$. This is a result of the error in approximating the operators $X(\hat{\theta}_i)$ using a Riemannian sum with $N_H$ elements. Indeed, it is guaranteed that only in the limit $N_H \rightarrow \infty$ the approximate score in Method $1$ converges to the true optimal score. Finally, we observe that Method $3$ saturates around its optimal value already for $N_O=d_I \times d_O=4$ outcomes. This is expected since, for a MSE cost function, projective measurements are optimal~\cite{macieszczak2014bayesian}. In Fig.~\ref{fig::M3_differentNo_alltimes}, we focus on this point, by comparing Method $3$ for some fixed values of $N_O$ as a function of time $t$, for the whole interval of time evaluated in this problem. Here we can see that while there is an improvement in increasing $N_O$ up to $4$, the curve for $N_O=16$ lies on top of that of $N_O=4$, demonstrating that there is indeed no advantage in increasing the number of outcomes beyond $N_O=d_I \times d_O$.

Another interesting point to make here is that the score clearly depends on time of evolution. In particular, in Fig.~\ref{fig::M3_differentNo_alltimes} we observe that there exist times $t < \infty$ where the score is much better (recall that for a MSE cost function, the optimal score is being minimized) than at $t \rightarrow \infty$. This means that measuring the probe in the transient regime can be advantageous over estimation performed after reaching the steady state. Indeed, this effect has been observed in thermometry previously~\cite{PhysRevA.91.012331}.For the steady state, the optimal measurement strategy is known. In particular, in this case the auxiliary system is useless and the optimal measurement is a PVM in the basis of the probe Hamiltonian~\cite{PhysRevLett.114.220405}. 
However, as we discuss in the next paragraph, here we show that this is not the case for the transient regime, where entanglement with the probe leads to a more precise estimation. Our techniques therefore allow to determine the optimal probe and memory state, as well as the measurements and estimators in this difficult regime. 

\begin{figure}
\begin{center}
	\includegraphics[width=\columnwidth]{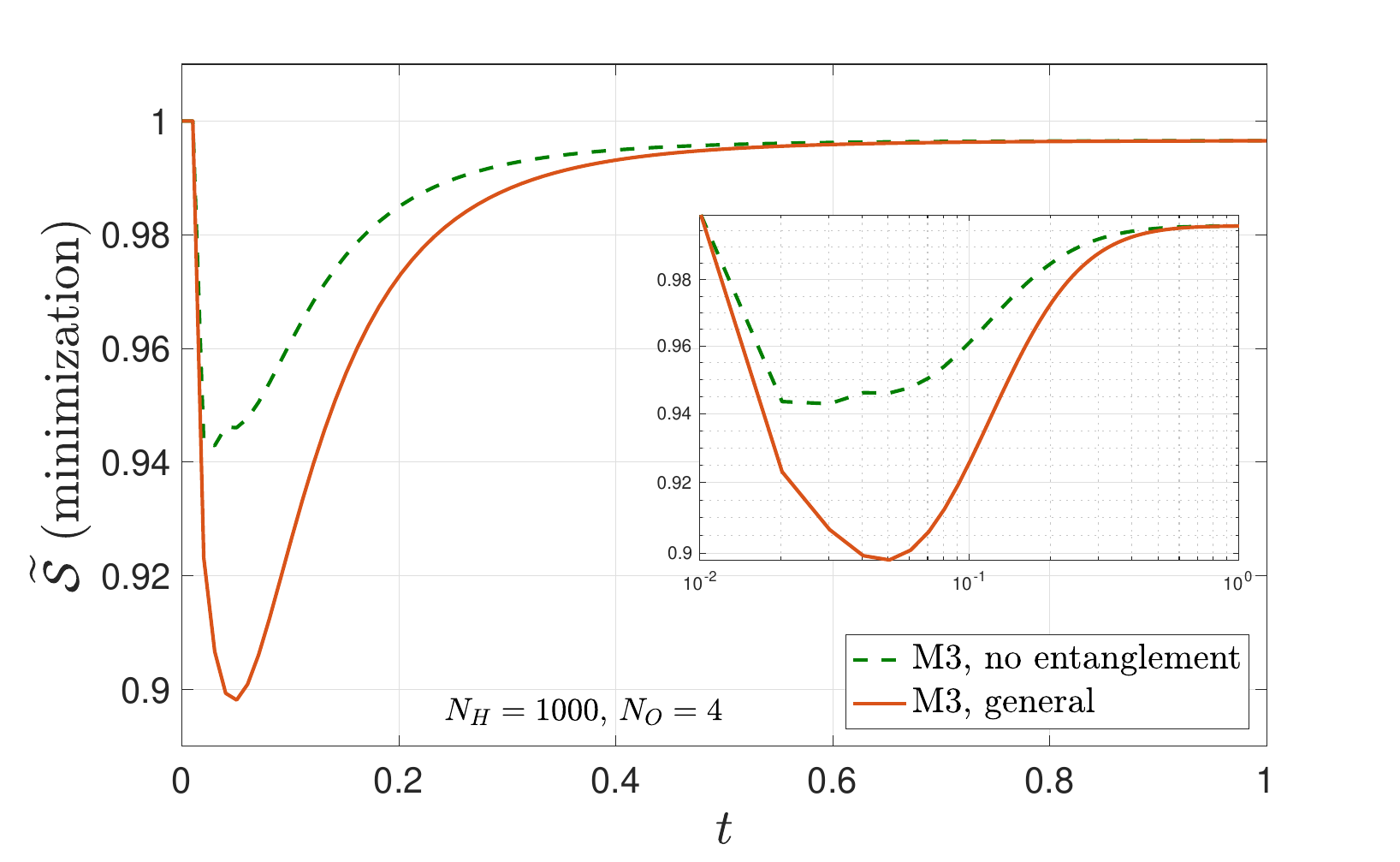}
    \caption{\textbf{Thermometry (Example 2): Advantage of entanglement in the transient regime.} The main panel shows the approximate score $\widetilde{\mathcal{S}}$ (computed via Method $3$) as a function of the evolution time $t$ for a fixed number of outcomes $N_O = 4$. The parameters chosen are as in Fig.~\ref{fig::allmethods_allNo_fixedtime}.  All values are renormalized by the maximum value of $\widetilde{\cS}$ in the plot. The inset shows the same curves plotted in a log-log scale. We see that there are times $t$ for which the precision of estimation is better than in the steady state $t \rightarrow \infty$. This can be understood as an advantage arising from having entanglement with the probe: the entanglement allows the transfer of the information about the parameter into the memory system which is itself not subject to the dephasing dynamics of the master equation. As a consequence, measuring the entangled probe and memory systems before the joint system thermalizes provides a significant advantage.}
    \label{fig::noentanglement}
\end{center}
\end{figure}

Finally, in order to investigate the role of entanglement in the transient regime of temperature estimation, we compare two different measurement strategies: a general strategy where the memory can be initially entangled with the probe, and the scenario where the initial state of the memory and probe is separable. Figure~\ref{fig::noentanglement} highlights the importance of the entangled auxiliary qubit. To this aim, we have focused again on Method 3, and depict the approximate score as a function of time, for strategies with and without entanglement. As one expects, at the limits of very short time or very long time the two kinds of strategies perform equally. While in the former this is because there has not been enough time to collect and add new information to the prior, in the latter case it is because after a long time, the system reaches a steady state regardless of the input state; namely, of it being entangled or not. However, at the transient regime, we observe that an auxiliary system entangled with the probe can significantly improve the score.

Let us emphasize that very often the parameter estimation problem described above cannot be solved analytically and is very difficult to solve numerically. In general, the effective evolution of the probe may result from a complicated master equation, which has to be evaluated many times. In our approach there is no need to evaluate the evolution for each potential state of the probe, as the only thing we need are the Choi operators associated with the effective channels. In this sense, our methods only require to solve the dynamics on a finite grid of parameter values, and thus makes finding the solution more tractable numerically.

\begin{figure}
\begin{center}
	\includegraphics[width=\columnwidth]{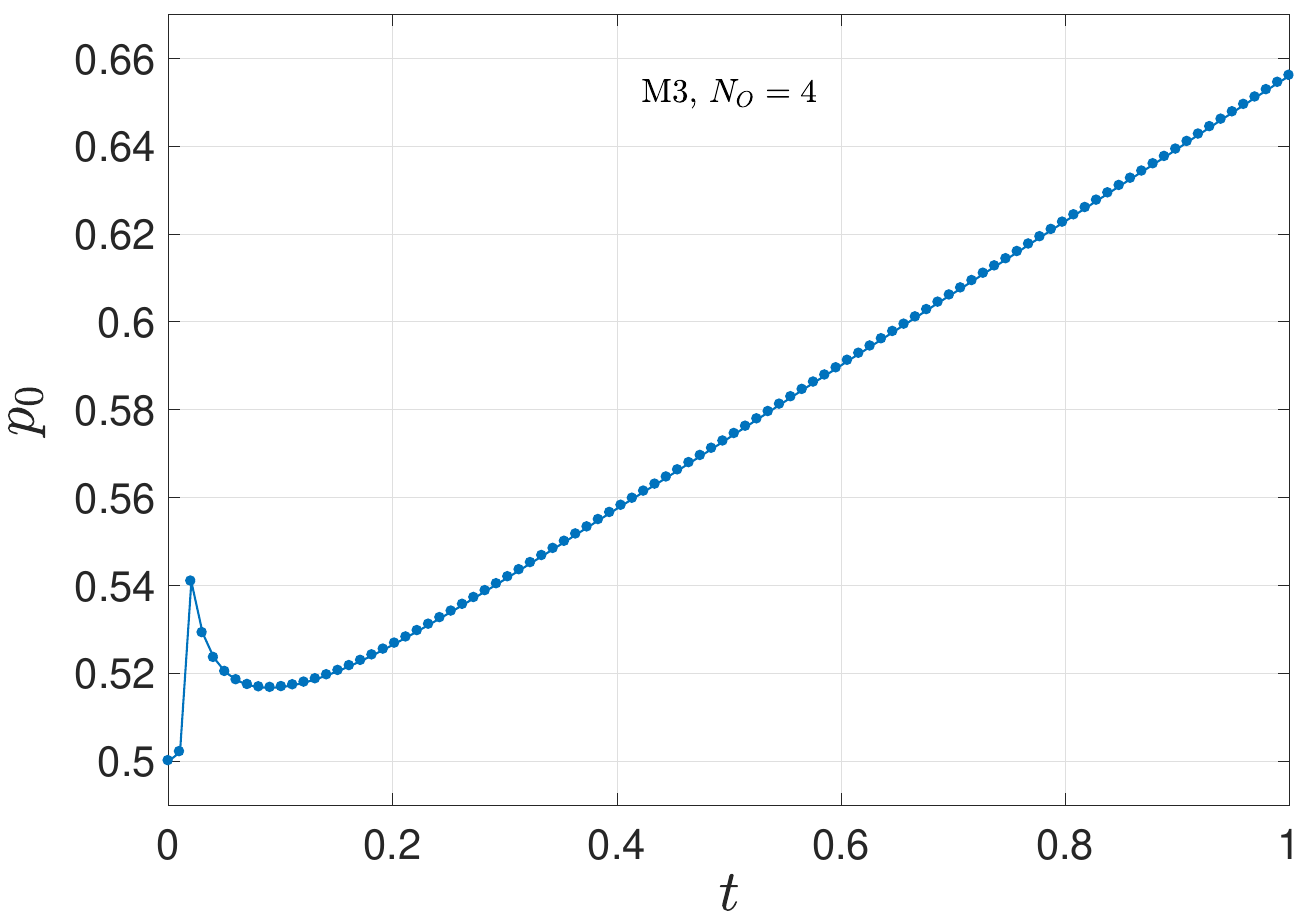}
    \caption{\textbf{Thermometry (Example 2): Optimal state.} The entanglement of the optimal initial state $\rho=\ketbra{\Psi}_{IA}$ in Eq.~\eqref{eq: intial state termo} as function of time found by method 3 for $N_O=4$. The corresponding score $\widetilde{\mathcal{S}}$ is given in Fig.~\ref{fig::M3_differentNo_alltimes}, the physical parameters are given in Fig.~\ref{fig::allmethods_allNo_fixedtime}.}
    \label{fig::schmidtcoeff}
\end{center}
\end{figure}

Finally, for the method 3 we investigate the optimal initial probe-ancilla state $\rho$ found by our algorithm for different times $t\in[0,1]$. We consider the case with four outcomes $N_O=4$, which is found to give the optimal precision. Here the tester has four elements $\{T_1,\dots,T_4\}$, and the initial state can be computed with the help of equations~(\ref{eq::tester2},\ref{eq::realization_rho}). Note that the state $\rho =\ketbra{\Psi}_{IA}$ is pure by construction, and its Schmidt diagonal form reads (for $p_0\geq p_1$)
\begin{equation}
\label{eq: intial state termo}
\ket{\Psi}_{IA} = \sqrt{p_0} \ket{\bm n}_I \ket{0}_A +\sqrt{p_1} \ket{-\bm n}_I \ket{1}_A, 
\end{equation}
where $\bm n$ is the Bloch vector corresponding to the state $\ketbra{\bm n} = \frac{1}{2}(\id+ \bm n^T \bm \sigma)$ and basis choice of the ancillary qubit plays no physical role. We find that for all times the optimal state is always Schmidt diagonal in the computational basis for the probe, and the Schmidt state corresponding the the larger value $p_0$ is always the ground state, i.e. $\ket{\bm n}_I=\ket{0}_I$. In contrast, the amount of entanglement in the optimal state depends on the interaction time $t$, as shown in Fig.~\ref{fig::schmidtcoeff}. In particular, for the two first times $t=0$ and $t=0.01$ the state is closed to be maximally entangled. At $t=0$ any state encodes no information on the parameter. At $t=0.01$ this is due to a numerical error, as the score $\widetilde S$ is still found to be maximal by the algorithm, see the orange line Fig.~ \ref{fig::M3_differentNo_alltimes}. Then we find that the entanglement in $\ket{\Psi}$, as captured by the value $p_0$ changes smoothly with $t$. Asymptotically $t\to \infty$,  we know that all initial states do equally well as the dynamics maps prepares as the steady state, which is product with the auxiliary system's state independent of the parameter. Moreover, from Fig.~\ref{fig::noentanglement} it is clear that the presence of entanglement in 
the initial state does not give any substantial improvement for $t$ close to one. We have also considered the optimal measurements $\{M_1,\dots, M_4\}$ in Eq.~\eqref{eq::realization_M} found by the algorithm. We see that the POVM elements are given by rank-1 projectors. The states corresponding to the projectors are also Schmidt diagonal in the computational basis (for the probe), and the amount of entanglement first increases and then decreases for $t\geq 0.06$ (similar to Fig.~\ref{fig::schmidtcoeff}). All the data about the optimal strategies found with our methods is available in our repository~\cite{github-metrology}.

Lastly, as we have readily pointed out our methods can be simply adopted to other reward functions. As an example, in the Appendix~\ref{app:proof_master_eq} we address the thermometry problem with the mean square logarithmic error as the reward function---which has gained attention in recent years due to respecting scale-invariance properties~\cite{Rubio2022,PhysRevLett.127.190402,PhysRevLett.128.130502}.
\subsection{Example 3: Multi-parameter estimation -- SU(2) gates}\label{sec::example_su2}

For our final example, we will consider a more complex metrology problem which involves multiple parameters. This is the problem of estimating any qubit unitary---the group SU(2). 

As a first observation, note that any qubit unitary operator can be parameterized in terms of three independent parameters $\theta \coloneqq (\theta^x, \theta^y, \theta^z)$, with $0 \leq \theta^i < 2 \pi$ for all $i\in\{x,y,z\}$, as
\begin{equation}
    U_{\theta} = e^{-i (\theta^x \sigma_x + \theta^y \sigma_y + \theta^z \sigma_z)}.
\end{equation}
Here, $\sigma_i$ for $i \in \{x, y, z\}$ are the three Pauli operators. Since these generators do not commute, the estimation of the unitary $U_{\theta}$ (or equivalently the parameter vector $\theta$) is a multiparameter estimation problem. The unitary channel that acts on the probe system and encodes the parameter $\theta$ is then given simply by $\Ecal_{\theta}[\cdot]=U_{\theta}(\cdot)U_{\theta}^\dagger$, and will have a Choi operator $C_\theta$ associated to it.

We take a natural reward function that captures how close the estimated unitary is from the actual one, which is the fidelity, i.e., 
\begin{align}\label{eq::cost_fidelity}
    r(\theta, \hat{\theta}_i) = \frac{1}{d^2}\tr(C_\theta\,C_{\hat{\theta}_i}),
\end{align}
where in this example $d=2$. Here $C_{\hat{\theta}_i}$---and for later reference $C_{\theta_k}$---are defined analogously to $C_{\theta}$, for a vector of estimator values $\hat{\theta}_i = (\hat{\theta}^x_{a},\hat{\theta}^y_{b},\hat{\theta}^z_{c})$ and for a vector of discretization values $\theta_k=(\theta^x_a,\theta^y_b,\theta^z_c)$.

Here, we again analyze the cases of two different prior distributions, a uniform prior, as in Eq.~\eqref{eq::uniformprior}, and a Gaussian prior, as in Eq.~\eqref{eq::gaussianprior}. 

The parameter vector $\theta=(\theta^x,\theta^y,\theta^z)$ is discretized into values $\theta_k=(\theta^x_a,\theta^y_b,\theta^z_c)$, where each of the three elements follow the discretization in Eq.~\eqref{eq::discretization}, with $\theta_\text{min}=-\pi$ and $\theta_\text{max}=\pi$, and all with the same value of $N_H=n_H$. Notice that this will amount to a final number of different discretization values of $N_H=n_H^3$.

The initial set of estimators $\hat{\theta}_i = (\hat{\theta}^x_{a},\hat{\theta}^y_{b},\hat{\theta}^z_{c})$ is also set according to Eq.~\eqref{eq::estimators} for each parameter estimator, using the same values of $\theta_\text{min}=-\pi$ and $\theta_\text{max}=\pi$, and all with the same value of $N_O=n_O$. Here again this amounts to a total number of outcomes equal to $N_O=n_O^3$, analogously to the indexation of the parameter discretization.

\begin{figure}
\begin{center}
	\includegraphics[width=\columnwidth]{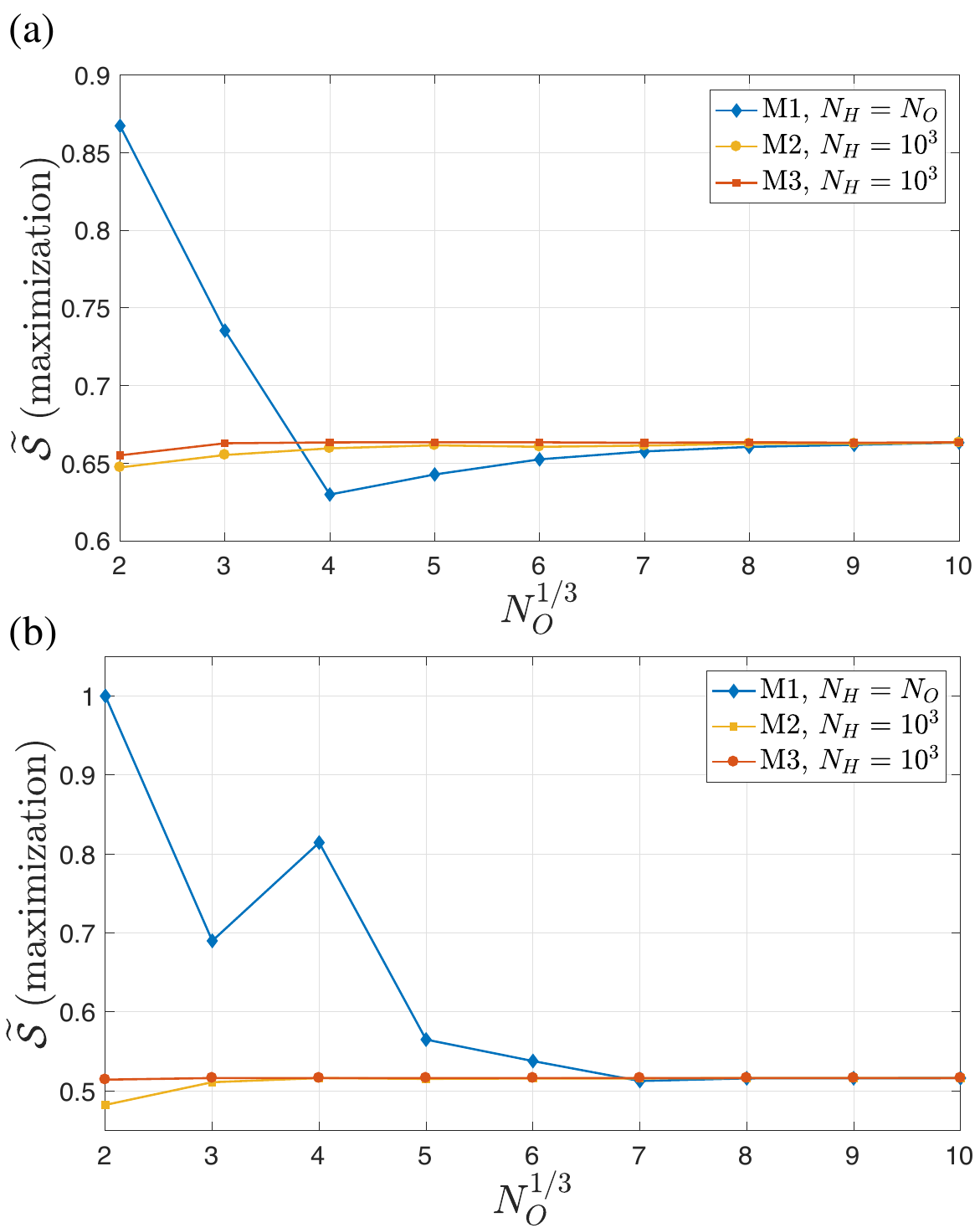}
    \caption{\textbf{SU($2$) estimation (Example 3).} The maximum approximate score $\widetilde{\mathcal{S}}$ for an SU(2) multiparameter estimation problem.  Both panels show the scores corresponding to Methods M1, M2, and M3 as a function of the cubic root of the total number of outcomes $N_O^{1/3}\in\{2,\ldots,10\}$, for different prior distributions of the phase parameters $(\theta^x,\theta^y,\theta^z)$. Panel $(a)$ shows the case of a uniform prior while $(b)$ corresponds to a Gaussian prior. Each of the three parameters ranges from $\theta_\text{min}=-\pi$ to $\theta_\text{max}=\pi$. The considered cost function is the fidelity in Eq.~\eqref{eq::cost_fidelity}.}
\label{fig::su2}
\end{center}
\end{figure}

We now discuss the application of our methods to this specific problem, plotting our results in Fig.~\ref{fig::su2}.

\textbf{Method 1.} To apply Method 1, we again set $n_H=n_O=n$ for each of the three parameters, amounting to a total number of discretization values and of outcomes equal to $N=n^3$. For the results presented here, we look at values of $n=N^{1/3}\in\{2,\ldots,10\}$. 

\textbf{Method 2.} Here we fix the number of outcomes $n_O=N_O^{1/3}\in\{2,\ldots,10\}$ and set the number of hypotheses to be $N_H = n_H^3 = 10^3$. Notice that while the total number of hypothesis is $1000$, each of the three parameters is discretized in only 10 different values. These values for the estimator are not expected to be optimal but are nevertheless used in Method 2, serving as an starting point for the estimator optimization in Method 3.

\textbf{Method 3.} To apply the seesaw in Method 3, we again begin with the solution provided by Method 2 for each $N_O$ as a starting point. In this case, we optimize over the estimators using standard gradient descent techniques. Therefore, the first step in the seesaw is the SDP in Method 2, while the second step is a heuristic search over the estimators for a fixed tester.

\textbf{Results.} In Fig.~\ref{fig::su2}, we plot the maximal approximate scores $\widetilde{S}$ obtained via the three methods outlined above. Panel (a) in Fig.~\ref{fig::su2} concerns the case of a uniform prior distribution and panel (b) that of a Gaussian prior distribution. In both cases we observe Method 3 converging to its final value of $\widetilde{S}$ with $N_O^{1/3}=3$, i.e. total number of outcomes $N_O=27$. This is consistent with the fact that, in this case, extremal testers have at most $(d_I\times d_O)^2=16$ outcomes, and hence, for $N_O^{1/3}=2$ (total number of outcomes $N_O=8$) we are not yet optimizing over all possible extremal testers. This is an interesting case where extremal non-projective POVMs with $(d_I\times d_O)^2$ outcomes show improvement over $(d_I\times d_O)$-outcome PVMs. Method 2 quickly approaches this same value, while Method 1 requires higher values of $N_O^{1/3}$. Nevertheless, as expected, for a larger number of outcomes, namely $N_O^{1/3}=10$, all methods yield the same result.

\section{Conclusion and Outlook}

We introduced a new set of tools for addressing Bayesian parameter estimation problems, applying techniques from the formalism of higher-order operations and drawing inspiration from the problem of channel discrimination. The key insight that we exploit consists of describing the quantum strategy, i.e., the state preparation and the measurement (see Fig.~\ref{fig:setup}), as a single operation called a quantum tester. The later is characterized by SDP constraints, and can thus be optimized efficiently.
We developed three methods for determining the state of the probe, the measurement, and the estimators in any parameter estimation problem, regardless of prior distribution, reward function, or description of the quantum evolution.

The first method exploits the connection between the Bayesian approach to parameter estimation and quantum channel discrimination. By discretizing the parameter to a finite set of values, and by furthermore associating each value of the estimator to a value of the discretized parameter, one can directly map a parameter estimation task onto a channel discrimination one, albeit with a reward function which inherits the geometry of the original parameter set. We leveraged this connection to create a general method for approximating the optimal solution of the estimation problem within any arbitrary precision. We also proved that our approximation converges to the optimal score. Although this method is conceptually simple and comes with a convergence guarantee, it may nonetheless be computationally demanding since the size of the optimization variables increases with the finess of parameter discretization.
Our second method computes an approximation of the optimal quantum strategy for a fixed set of estimators. This method is less computationally demanding in general, but it relies on a good guess for optimal estimators, which is not always available.
To address this drawback, our third method iteratively combines an optimization over the quantum strategy and over the estimators, and hence does not require any previous knowledge over the estimator.

A key advantage of our methods is their universal applicability. They can be used for any parameter estimation problem, regardless of the nature of parameter-encoding or the number of parameters to be estimated. To showcase this wide-ranging applicability, we examined three distinct case studies of high practical importance: local phase estimation, thermometry, and SU(2) estimation.

We also developed tools to bound the performance of estimation strategies that do not require entanglement and used them to show that, in the thermometry problem, probe states that are entangled with an auxiliary system lead to a more precise estimation of the temperature parameter, particularly at finite times.

Our work provides a starting platform for the application of higher-order operations to the problem of Bayesian parameter estimation. We conclude by summarizing further research directions that could draw further benefits from this approach. 

\textit{Generalization to many-shots.}
The quantum strategies explored here concern a situation in which, at each independent realization of the experiment, one is given access to a single call, or copy, of the channel that encodes the parameter ${\theta}$. It is also the case that, in more general scenarios, where one has access to multiple calls at once, the optimization of the quantum strategy can be done with SDP as well. Multiple-copy testers can take different forms, describing different classes of quantum strategies, such as parallel (non-adaptive) and sequential (adaptive), or even those involving an indefinite causal order. Such testers have been defined in Refs.~\cite{bavaresco21,bavaresco22} and explored in a frequentist approach to metrology~\cite{liu2023}. These techniques can also be applied to multiple-copy Bayesian estimation protocols, and be exploited to investigate, for example, whether different classes of estimation strategies can lead to higher precision in the parameter estimation. Similarly, strategies with an indefinite causal-order could lead to establishing new types of metrological resources.

While in the multi-shot scenario one can find the global optimal protocol as explained above, it can be hard to implement in practice, due to the exponential growth of of Hilbert space dimension. Alternatively, one can seek greedy optimal algorithms~\cite{papadimitriou1998combinatorial}. In such a strategy, one would (i) perform the optimisation protocol as subscribed in our work, (ii) update the prior distribution to the posterior distribution based on the outcome and repeat steps (i) and (ii) until all shots are consumed. Despite not being necessarily globally optimal, this strategy can be very strong and has shown to be asymptotically optimal in some cases~\cite{Feige1998}. An example of our interest is in Bayesian equilibrium thermometry~\cite{PhysRevLett.128.130502}. Regardless of global optimality, it is practically easier to implement such multi-shot strategies since the required operations do not involve exponentially increasing Hilbert spaces. 

\textit{Applications to complex noise models.} 
One of the key advantages of our approach is its versatility in handling various types of parameter-encoding dynamics. Often sensing methods are limited to specific parameter-encoding channels, however, our approach can effectively model and accommodate any type of dynamics and address different types of noise. Specifically, if one has a good description of the noise appearing in the measurement process, one can simply incorporate this noise into the encoding channel and compute optimal testers and optimal estimators according to any one of the three methods. Therefore, applying these techniques to real (noisy) experimental settings to infer their actual performance would be an interesting next step. 

\textit{Quantum metrology techniques for asymptotic quantum channel discrimination.} 
In this study, we utilized higher-order operations, a technique previously used to study channel discrimination, to provide a nearly optimal solution for the quantum parameter estimation problem. It would be interesting to explore whether the reverse approach could also yield novel insights into the field of channel discrimination. Specifically, one could investigate whether leveraging asymptotic theoretical results from quantum metrology, such as the Heisenberg scaling, can contribute to the investigation of asymptotic quantum channel discrimination. This direction holds promise for gaining a deeper understanding of the relationship between channel discrimination and quantum metrology.

\textit{Connections with the multi-hypothesis testing problem.} The discretization of the parameter space that we perform  suggests that the Bayesian estimation problem can be connected with a multi-hypothesis testing problem~\cite{rupert2012simultaneous}. However, it should be noted that this connection is only partial. Indeed, our work exploits the fact that Bayesian estimation can be seen as a multi-hypothesis testing problem  with ($i$) a continuous set of hypotheses and with ($ii$) a specific geometry on the ``hypothesis space'' as captured by the cost function. Still, we believe that the methods developed here (Method $1$ in particular) could be potentially useful for determining bounds on the errors for the multi-hypothesis testing problem. Another interesting question is whether some of the bounds on error probabilities arising in the multi-hypothesis testing scenario could be also applicable in the Bayesian setting.
\\

All code developed for this work is freely available in our online repository~\cite{github-metrology}.

\section*{Acknowledgments}
We are thankful to Mart\'i Perarnau-Llobet, Nicolas Brunner,  Sumeet Khatri, and Rafal Demkowicz-Dobrzanski for fruitful discussions. J.B., P.L.B., and P.S. acknowledge funding from the Swiss National Science Foundation (SNSF) through the funding schemes SPF and NCCR SwissMAP. The authors acknowledge TU Wien Bibliothek for financial support through its Open Access Funding Programme. This research was funded in part by the Austrian Science Fund (FWF) [Grant No. I 6047-N].

\onecolumngrid
\appendix

\section*{APPENDIX}

\section{Approximations for strategies without entanglement}\label{app::noentanglement}

If the particular problem at hand does not allow for the use of auxiliary systems that may be entangled with the target system, the estimation problem in Eq.~\eqref{eq::prob_map} above reduces to the following:

\begin{equation}
    p(i|\theta) = \tr\left(\Ecal_{\theta}[\rho]\,M_i\right),
\end{equation}
where now $\rho\in\Lcal(\Hcal^I)$ and $M_i\in\Lcal(\Hcal^O)$, or, in the Choi representation,
\begin{equation}
    p(i|\theta) = \tr\Big(
    C_{\theta}\,(\rho^{T}\otimes M_i)\Big).
\end{equation}

In this case, the optimization over the quantum strategy (state and measurement) of any linear function of $p(i|\theta)$ is no longer an SDP. The tester in this case, which does not require an auxiliary system for implementation, is given by
\begin{equation}\label{eq::tester_noent}
    T_i = \rho^{T}\otimes M_i,
\end{equation}
where $\rho\in\Hcal^I$ and $M_i\in\Hcal^O$. Unfortunately, this kind of tester does not admit a simple mathematical characterization. In fact, the problem of characterizing this kind of tester is mathematically very similar to the problem of characterizing separable states, except in this case we are interested in the `separability' between the operators acting on the input and output spaces of the tester.

For this reason, techniques used to approximate the set of separable states can be applied here to approximate the set of testers that do not require an auxiliary system, or entanglement, for its implementation. 

Outer approximations of the set of testers that do not require entanglement are useful to determine whether or not entanglement/memory is advantageous in an estimation task---if the score achieved by a general tester is better than the best score achieved by a tester in the outer approximation, then entanglement is useful. On the other hand, inner approximations are useful to determine a lower bound on how well a tester without entanglement can perform in a given task. 

An example of outer approximation is the PPT condition~\cite{peres96}. The set of all testers whose elements have a positive partial transpose, i.e., the set of all $T=\{T_i\}$ that satisfy
\begin{align}
    T_i &\geq 0 \ \ \ \forall \, i \\
    \sum_i T_i &= \sigma\otimes\id^{O} \\
    (T_i)^{\text{T}_I} &\geq 0 \ \ \ \forall \, i 
\end{align}
is an outer approximation of the set of testers that satisfy Eq.~\eqref{eq::tester_noent}. Similarly, the imposition of the condition of k-symmetric extensibility~\cite{doherty02} on the tester elements defines an outer approximation for the set of testers that require entanglement.  Any other SDP method of approximating the set of separable states can also be applied to approximate the set of testers without entanglement, although the resulting approximation may not be tight or may not converge to the actual set.

Inner approximations can, for example, be achieved by applying a see-saw method of optimizing over states and measurements separately over many iterations. This method is not guaranteed to converge, but will yield a bound. 

\section{Convergence of the approximations}\label{app:convergence}

The cornerstone of our results is the discretization of the estimators and the hypotheses. The intuition suggests that as the discretization is made finer, the approximation becomes more precise and converges to the exact value. Here, we make this statement more rigorous. We focus on a score function that needs to be maximised; for those that require minimization a similar argument holds.

First, let us denote the optimal protocol by $\{\{T_i^*\}, \{\hat{ \theta}_i^*\}\}$, it maximizes the score in Eq.~\eqref{eq:exact_score_pair} to it's optimal value $\cS^*$. We know that the optimal protocol has at most $D\coloneqq (d_I\times d_O)^2$ elements, therefore
\begin{align}
    \{\{T_i^*\}, \{\hat{ \theta}_i^*\}\} &\coloneqq 
    \underset{\{T_i\}, \{{\hat \theta}_i\}}{\arg\max} 
    \sum_{i=1}^{D} \tr \left({X}(\hat{\theta}_i)\,T_i\right)\\
    \cS^* &\coloneqq  \underset{\{T_i\}, \{{\hat \theta}_i\}}{\max} 
    \sum_{i=1}^{D} \tr \left({X}(\hat{\theta}_i)\,T_i\right) =\sum_{i=1}^{D} \tr \left({X}(\hat{\theta}_i^*)\,T_i^*\right),
\end{align}

Now consider another protocol in which the estimators are fixed to $\{\hat{\theta}_i\}_{i=1}^{N_O}$, and the tester $\{\tilde T_i\}_{i=1}^{N_O}$ is the solution of the SDP maximizing the score for the given estimators 
\begin{equation}
    \{\tilde T_i\} =  \underset{\{T_i\}}{\arg\max}\sum_{i=1}^{N_O} \tr \left({X}(\hat{\theta}_i)\,T_i\right).
\end{equation}
The protocol achieves a certain score
\begin{equation}
    \widetilde{\cS}^* = \underset{\{T_i\}}{\max}\sum_{i=1}^{N_O} \tr \left({X}(\hat{\theta}_i)\,T_i\right) = \sum_{i=1}^{N_O} \tr \left({X}(\hat{\theta}_i)\,\tilde T_i\right)\leq \cS^*.
\end{equation}

Next, let us consider the two sets of estimators $\{\hat \theta_i^*\}_{i=1}^{D}$ and $\{\hat \theta_i\}_{i=1}^{N_O}$ and for each $\hat \theta_i^*$ define $\bar \theta_i^*$ to be the closest value in the second set, i.e.
\begin{align}
    {\bar \theta}_i^* &\coloneqq \underset{\hat{\theta}_k|1\leq k \leq N_O }{\arg\min}|\hat{\theta}_k-\hat{\theta}_i^*|\\
    \epsilon_i &\coloneqq \hat{\theta}_i^* - {\bar \theta}_i^*,\label{eq:epsilon_i}
\end{align}
where $\epsilon_i$ quantifies how different these values are. For simplicity we also introduce $\epsilon \coloneqq \max_i |\epsilon_i|$. Note that for concreteness we here used the absolute value of the difference $|\hat{\theta}_i-\hat{\theta}_k^*|$ as a distance between the estimated values, however any other distance $d(\hat{\theta}_i,\hat{\theta}_k^*)$ could be used here and below to define $\bar \theta^*_k$ and $\epsilon$ instead (e.g. in the multiparameter case).
The new values $\{\bar \theta_i^*\}$ allow us to define the protocol $\{\{T_i^*\},\{\bar \theta_i^*\}\}$, where the tester are taken from the optimal protocol but the estimators have been modified. It achieves a certain score
\begin{equation}
     \sum_{i=1}^{D}  \tr \left({X}(\bar{\theta}_i^*)\,T_i^*\right) \leq \max_{\{T_i\}_{i=1}^{D}}
    \sum_{i=1}^{D}  \tr \left({X}(\bar{\theta}_i^*)\,T_i\right)\leq \underset{\{T_i\}_{i=1}^{N_O}}{\max}\sum_{i=1}^{N_O} \tr \left({X}(\hat{\theta}_i)\,T_i\right) = \widetilde{\cS}^*.
\end{equation}
Here, we used the fact that by construction $\{\bar \theta_k^*\}_{i=1}^{D}$ form a subset of $\{\hat \theta_i\}_{i=1}^{N_O}$, therefore the maximization over the tester $\{T_i\}_{i=1}^{N_O}$ includes the maximization over the tester $\{T_i\}_{i=1}^{D}$ (some of the tester elements can be identically zero). 

Our next goal is to bound the deviation between $\sum_{i=1}^{D}  \tr \left({X}(\bar{\theta}_i^*)\,T_i^*\right)$ and the optimal score $\cS^*$, which are obtained with the same tester. To do so we recall their Bayesian interpretation in terms of posterior parameter distribution in Eq.~\eqref{eq::continuous_score}
\begin{align}
    \sum_{i=1}^{D}  \tr \left({X}(\hat {\theta}_i^*)\,T_i^*\right) = \sum_{i=1}^{D} p(i) \int \dd \theta \, p(\theta|i) r(\theta,\hat \theta_i^*) =  \sum_{i=1}^{D} p(i) \, \mathds{E}^{(i)}[r(\theta,\hat \theta_i^*)]
\end{align}
where each expected values $\mathds{E}^{(i)}[\cdot]$ is taken with respect to the probability distribution $p(\theta|i)$. This allows us to write 
\begin{equation}\label{eq: bound difference scores}
    \cS^* - \widetilde{\cS}^* \leq \cS^* - \sum_{i=1}^{D}  \tr \left({X}(\bar{\theta}_i^*)\,T_i^*\right) =     \sum_{i=1}^{D} p(i) \, \big(\mathds{E}^{(i)}[r(\theta,\hat \theta_i^*)] -\mathds{E}^{(i)}[r(\theta,\bar \theta_i^*)]\big) = \sum_{i=1}^{D} p(i) \mathds{E}^{(i)}[r(\theta,\hat \theta_i^*)-r(\theta,\bar \theta_i^*)].
\end{equation}
Here, it is intuitively clear that for nearby value $\hat \theta_i^* $ and $\bar \theta_i^*$ the expected values $\mathds{E}^{(i)}[r(\theta,\hat \theta_i^*)]$ and $\mathds{E}^{(i)}[r(\theta,\bar\theta_i^*)]$ will also be close, provided that the reward function $r$ is regular enough. For simplicity let us now assume that it is Lipschitz continuous, i.e.,
\begin{equation}
    |r(\theta,\hat \theta_i^*)-r(\theta,\bar \theta_i^*)| \leq K_r\,  \epsilon \qquad \text{if} \qquad 
|\hat{\theta}_i^* - {\bar \theta}_i^*| \leq \epsilon,
\end{equation}
for any small enough $\epsilon\leq \delta$ (here $K_r$ might depend on delta)
which directly implies $  \widetilde{\cS}^* \geq \cS^* - K_r \, \epsilon$.  Finally, for a scalar parameter the $N_O$ estimators $\{\hat \theta_i\}$ can be chosen such that $\epsilon\leq \frac{L}{N_O}$, where $L$ is some constant depending on the prior. This would guarantee the convergence to the optimal score with
\begin{equation}
    0\leq \cS^* -  \widetilde{\cS}^* \leq \frac{K_r L}{N_O}.
\end{equation}

\subsection{The case of reward functions that are not  Lipschitz continuous}

Notably, Lipschitz continuity of the reward function is not necessary to guarantee the convergence of the score $ \widetilde{\cS}^* \to \cS^*$. However, in such cases it seems difficult to make a general statement, which might furthermore require to assume some regularity of the prior. Nevertheless, for illustration let us consider a piece-wise constant reward function 
\begin{equation}
    r(\theta,\hat\theta) = \begin{cases}
    1 & |\theta-\hat \theta|\leq \Delta \\
    0 & \text{otherwise},
    \end{cases}
\end{equation}
that can be used to define a confidence interval for the parameter. This reward function coincides with the recent proposal in Ref.~\cite{meyer2023quantum}. This function is manifestly discontinuous, with $r(\theta,\hat \theta_i^*)- r(\theta ,\bar \theta_i^*)$ taking the value $+1$ on an interval $\theta\in I_+^i$ of width $|\hat \theta^*_i-\bar \theta^*_i|\leq \epsilon$, the value $-1$ on another interval of the same width, and is otherwise zero. From Eq.~\eqref{eq: bound difference scores} we then find
\begin{equation}
    \cS^*-\widetilde{\cS}^*\leq \sum_{i=1}^{D} p(i) \mathds{E}^{(i)}[r(\theta,\hat \theta_i^*)-r(\theta,\bar \theta_i^*)] \leq  \sum_{i=1}^{D} p(i) \text{Pr}^{(i)}[\theta\in I_+^i]
\end{equation}
where each probability $\text{Pr}^{(i)}[f(\theta)]=\int \dd \theta f(\theta) p(\theta|i)$ is taken over the conditional distribution $p(\theta|i)$. Defining the  union of all the intervals $I_+ = \cup_{i=1}^{D} I_+^i$ we can further upper bound the score difference with
\begin{equation}
    \cS^*-\cS\leq  \sum_{i=1}^{D} p(i) \text{Pr}^{(i)}[\theta\in I_+^i] \leq  \sum_{i=1}^{D} p(i) \text{Pr}^{(i)}[\theta\in I_+] = \text{Pr}[\theta\in I_+], 
\end{equation}
where in the last term the probability is taken over the prior distribution $p(\theta) = \sum_i p(i) p(\theta|i)$, and we used
\begin{equation}
   \sum_{i=1}^{D} p(i) \text{Pr}^{(i)}[\theta\in I_+] =   \int \dd \theta \sum_{i=1}^{D} p(i) p(\theta|i) \chi_{I_+}(\theta) =   \int \dd \theta  p(\theta) \chi_{I_+}(\theta)
\end{equation}
for the indicator function $\chi_{I_+}=\begin{cases} 1 & \theta \in I_+ \\ 0 & \theta \notin I_+ \end{cases}$. 
Here, the set $I_+$ is of measure at most $D\epsilon$. Thus, for any regular prior $p(\theta)$ supported ($p(\theta)\neq 0$) on a set of finite measure, the probability $\text{Pr}[\theta\in I_+]$ to find a value inside $I_+$ converges to zero with $\epsilon$. In particular, for any prior with a bounded density $p(\theta)\leq p_*$ we find
\begin{equation}
    \cS^*-\cS\leq\text{Pr}[\theta\in I_+] =\int \dd \theta p(\theta) \chi_{I_+}(\theta)\leq \int \dd \theta p_* \chi_{I_+}(\theta)\leq  D \epsilon \, p_*.
\end{equation}

\subsection{Convergence for the MSE}

As a first example, let us have a look into the MSE that we used for the thermometry problem.

In this case the score function is a cost which has to be minimized, so to match to the notation with the previous section we consider maximization of $r(\theta,\hat \theta)= -(\theta -\hat \theta)^2$. We have
\begin{align}
    r(\theta,\hat \theta_i^*)-r(\theta,\bar \theta_i^*)& = -(\theta - \hat \theta_i^*)^2 + -(\theta - \bar \theta_i^*)^2
    \\
    &= -(\theta - \hat \theta_i^*)^2 + (\theta - \hat \theta_i^*+\epsilon_i)^2\\
    & = \epsilon_i^2 + 2\epsilon_i (\theta-\hat \theta_i^*),
\end{align}
where $\epsilon_i$ is defined in Eq.~\eqref{eq:epsilon_i}. Plugging this in the Eq.~\eqref{eq: bound difference scores} one gets
\begin{equation}
\cS^* - \widetilde{\cS}^*  \leq  \sum_{i=1}^{D} p(i) \mathds{E}^{(i)}[r(\theta,\hat \theta_i^*)-r(\theta,\bar \theta_i^*)] =  \sum_{i=1}^{D} p(i) \left( \epsilon_i^2 +2 \epsilon_i \mathds{E}^{(i)}[(\theta-\hat \theta_i^*)]\right).
\end{equation}
But for the MSE we know that the optimal estimator is the mean, i.e. $\hat\theta_i^* =  \mathds{E}^{(i)}[\theta] = \int \dd\theta \, \theta p(\theta|i)$. Therefore, the second term is zero and we find 
\begin{equation}
\cS^* - \widetilde{\cS}^*  \leq  \sum_{i=1}^{D} p(i)  \epsilon_i^2 \leq \epsilon^2.
\end{equation}
with $\epsilon=\max_i |\epsilon_i|$.

\subsection{Convergence for the $\cos^2$ reward function}

In the phase estimation problem we considered a reward function that reads $r(\theta,\hat{\theta}_i^*) = \cos^2\left(\frac{\theta-\hat{\theta}_i^*}{2}\right)$. 
First of all, note that in this case the optimal estimator can be find in a closed form. To do so, we first rewrite the score for the posterior distribution $p(\theta|i)$ as
\begin{equation}\label{eq app: cosine cost}     
    \mathds{E}^{(i)}\left[\cos^2\left(\frac{\theta-\hat{\theta}_i^*}{2}\right)\right]  
    =\mathds{E}^{(i)}\left[\frac{\cos(\theta-\hat{\theta}_i^*)+1}{2} \right]= \frac{1+\mathds{E}^{(i)}[\cos(\theta)]\cos(\hat \theta_i^*) + \mathds{E}^{(i)}[\sin(\theta)]\sin(\hat \theta_i^*) }{2}.
\end{equation}
Impose that the derivative of the score with respect to the estimator $\hat \theta_i^*$ is zero 
\begin{equation}\label{eq:opt_crit_cos2}
    \frac{\partial}{\partial \hat \theta_i^*} \mathds{E}^{(i)}\left[\cos^2\left(\frac{\theta-\hat{\theta}_i^*}{2}\right)\right] = 
    - \frac{1}{2}    \mathds{E}^{(i)}\left[\sin(\theta-\hat\theta_i^*)\right]
    =\frac{1}{2}( 
    -\mathds{E}^{(i)}[\cos(\theta)]\sin(\hat \theta_i^*) + \mathds{E}^{(i)}[\sin(\theta)]\cos(\hat \theta_i^*) 
    )=  0.
\end{equation}
Which is equivalent to 
\begin{equation}
    \tan(\hat \theta_i^*) = \frac{\mathds{E}^{(i)}[\sin(\theta)]}{\mathds{E}^{(i)}[\cos(\theta)]}
\end{equation}
and admits two solutions
\begin{align}
    \hat \theta_i^* = \arctan\left(
    \frac{\avg{\sin(\theta)}^{(i)}}{\avg{\cos(\theta)}^{(i)}}\right) \quad \text{or} \quad \hat \theta_i^* = \arctan\left(
    \frac{\avg{\sin(\theta)}^{(i)}}{\avg{\cos(\theta)}^{(i)}}
    \right)+\pi,
\end{align}
with the notation from the main text.  We then need to pick the value which gives the highest contribution to the reward in Eq.~\eqref{eq app: cosine cost}. In fact, up to a constant the reward is the scalar product between the vectors $(\avg{\cos(\theta)}^{(i)},\avg{\sin(\theta)}^{(i)})$ and $(\cos(\hat \theta^*_i),\sin(\hat \theta^*_i))$, so it's maximum is attained when the two vectors are in the same half of the disc. Since the range of arctan $\in [-\frac{\pi}{2},\frac{\pi}{2}]$ corresponds to positive cosine, the choice of the optimal estimator solution depends on the sign of $\avg{\cos(\theta)}^{(i)}$ 
\begin{equation}
     \hat \theta_i^* = \begin{cases} 
     \arctan\left( \frac{\avg{\sin(\theta)}^{(i)}}{\avg{\cos(\theta)}^{(i)}}\right) & \avg{\cos(\theta)}^{(i)}\geq 0,\\
      \arctan\left(
    \frac{\avg{\sin(\theta)}^{(i)}}{\avg{\cos(\theta)}^{(i)}}\right)+\pi & \text{otherwise}.
    \end{cases}
\end{equation}

Therefore, we can see that
\begin{align}
    \cS^* - \widetilde{\cS}^* & \leq  \sum_{i=1}^{D} p(i) \mathds{E}^{(i)}[r(\theta,\hat \theta_i^*)-r(\theta,\bar \theta_i^*)] \nonumber\\
    & = \sum_{i=1}^{D} p(i) \mathds{E}^{(i)}\left[\cos^2\left(\frac{\theta -\hat \theta_i^*}{2}\right)-\cos^2\left(\frac{\theta -\hat \theta_i^*}{2}\right)\cos^2(\frac{\epsilon_i}{2}) + \frac{1}{2} \sin(\theta -\hat \theta_i^*)\sin (\epsilon_i) - \sin^2\left(\frac{\theta -\hat \theta_i^*}{2}\right)\sin^2(\frac{\epsilon_i}{2})
    \right]\nonumber\\
    & = \sum_{i=1}^{D} p(i) \mathds{E}^{(i)}\left[\cos(\theta -\hat \theta_i^*)\sin^2(\frac{\epsilon_i}{2}) + \frac{1}{2} \sin(\theta -\hat \theta_i^*)\sin (\epsilon_i)
    \right]\nonumber\\
    & = \sum_{i=1}^{D} p(i) \mathds{E}^{(i)}\left[\cos(\theta -\hat \theta_i^*)\sin^2(\frac{\epsilon_i}{2})\right] + \frac{1}{2}\sum_{i=1}^{D} p(i) \mathds{E}^{(i)}\left[ \sin(\theta -\hat \theta_i^*)\sin (\epsilon_i)
    \right]\nonumber\\
    & =
    \sum_{i=1}^{D} p(i) \mathds{E}^{(i)}\left[\cos(\theta -\hat \theta_i^*)\sin^2(\frac{\epsilon_i}{2})\right]
    \leq
    \frac{1}{4}\sum_{i=1}^{D} p(i) \epsilon_i^2\mathds{E}^{(i)}\left[\cos(2(\theta -\hat \theta_i^*))\right]
    \leq  \frac{1}{4}\sum_{i=1}^{D} p(i) \epsilon_i^2 \leq \frac{\epsilon^2}{4},
\end{align}
where in the penultimate line we use the optimality criterion Eq.~\eqref{eq:opt_crit_cos2}, and used the definition of $\epsilon_i$ in Eq.~\eqref{eq:epsilon_i}, and $\epsilon=\max_i|\epsilon_i|$.
\section{Extension to multiparameter estimation}\label{app::multiparameter}

Take the two-parameter estimation example, where the parameters ${{\theta}^1}$ and ${{\theta}^2}$
are encoded via a channel $C_{{{\theta}^1},{{\theta}^2}}$. The estimates of these parameters are given respectively by $\{\hat{\theta}^1_{i_1}\}_{i_1=1}^{N_{O_1}}$ and $\{\hat{\theta}^2_{i_2}\}_{i_2=1}^{N_{O_2}}$. The continuous version of the problem can be expressed as 
\begin{equation}
    \cS = \sum_{i_1=1}^{N_{O_1}} \sum_{i_2=1}^{N_{O_2}}  \int d{{\theta}^1}\int d{{\theta}^2} \ p({{\theta}^1},{{\theta}^2}) \, r({{\theta}^1},{{\theta}^2},\hat{\theta}^1_{i_1},\hat{\theta}^2_{i_1}) \, \tr(C_{{{\theta}^1},{{\theta}^2}}\,T_{i_1,i_2}).
\end{equation}

We then choose a discretization of ${{\theta}^1}$ given by $\{\theta^1_{k_1}\}_{k_1=1}^{N_{H_1}}$ and of ${{\theta}^2}$ given by $\{\theta^2_{k_2}\}_{k_2=1}^{N_{H_2}}$, leading to the approximation
\begin{equation}
    \widetilde{\cS} = \sum_{i_1=1}^{N_{O_1}} \sum_{i_2=1}^{N_{O_2}} \sum_{k_1=1}^{N_{H_1}} \sum_{k_2=1}^{N_{H_2}} p(\theta^1_{k_1},\theta^2_{k_2}) \, r(\theta^1_{k_1},\theta^2_{k_2},\hat{\theta}^1_{i_1},\hat{\theta}^2_{i_2}) \, \tr(C_{\theta^1_{k_1},\theta^2_{k_2}}\,T_{i_1,i_2}).
\end{equation}
By mapping the indexes $(i_1,i_2)\mapsto i$, where $i\in\{1,\ldots,N_O\}$, $N_O=N_{O_1}N_{O_2}$, and 
$(k_1,k_2)\mapsto k$, where $k\in\{1,\ldots,N_H\}$, $N_H=N_{H_1}N_{H_2}$, and furthermore defining $\vec{\theta}_{k}\coloneqq (\theta^1_{k_1},\theta^2_{k_2})$ and $\vec{\hat{\theta}}_{i}\coloneqq (\hat{\theta}^1_{i_1},\hat{\theta}^2_{i_2})$, $S$ can be rewritten as
\begin{equation}
    \widetilde{\cS} = \sum_{i=1}^{N_O}  \sum_{k=1}^{N_H} p(\vec{\theta}_k) \, r(\vec{\theta}_k, \vec{\hat{\theta}}_i) \, \tr(C_{\vec{\theta}_k}\,T_i),
\end{equation}
which is equivalent to the single-parameter problem.

Hence, the techniques presented here are applicable to multiparameter estimation problems as well.

\section{Details of thermometry (Example 3)}\label{app:proof_master_eq}
Here, we provide some technical details for the Example B where we want to estimate the temperature of a bosonic bath. A qubit that is initially prepared in the state $\rho^p(0)=\big(\begin{smallmatrix}
  \rho_{11} & \rho_{12}\\
  \rho_{21} & 1-\rho_{11}
\end{smallmatrix}\big)$. Under the evolution~\eqref{eq:master_equation} the probe evolves to 
\begin{align}\label{eq:qubit_thermal_channel}
\begin{array}{l}
    {\cal E}_{\theta}(t)[\rho^{p}(0)] = \rho^p_{\theta}(t) = \left(\begin{array}{cc}
    \frac{{\mathrm{e}}^{-\Gamma t } \,{\left(\rho_{11}(2N_{B/F}+1) +  ({\mathrm{e}}^{\Gamma t }-1)(N_{B/F}+1)\right)}}{2\,N_{B/F}+1} & \rho_{12} \,{\mathrm{e}}^{-\frac{\Gamma t + 2i \epsilon t}{2}} \\
    \rho_{21} \,{\mathrm{e}}^{-\frac{\Gamma t - 2i \epsilon t}{2}} & \frac{{\mathrm{e}}^{-\Gamma t } \,\left( -\rho_{11}(2N_{B/F}+1) + N_{B/F}(1+{\mathrm{e}}^{\Gamma t }) + 1\right)}{2\,N_{B/F}+1}
\end{array}\right),\\
\mathrm{}\\
\textrm{where}\\
\mathrm{}\\
\;\;\Gamma = \Gamma_{\rm in} + \Gamma_{\rm out} = J(\epsilon) \,{\left(2\,N_{B/F}+1\right)}.
\end{array}
\end{align}
The Choi operator of the thermalization channel is then given by
\begin{align}
    \begin{array}{l}
    \sum_{ij} \Ecal_{\theta}(t)\left[\ketbra{i}{j}\right]\ot \ketbra{i}{j}=
    \left(\begin{array}{cccc}
    \frac{N_{B/F}+N_{B/F}\,{\mathrm{e}}^{-\Gamma t } +1}{{\left(2\,N_{B/F}+1\right)} } & 0 & 0 & {{\mathrm{e}}^{-\frac{\Gamma t + 2i \epsilon t}{2}}}\\
    0 & \frac{{\mathrm{e}}^{-\Gamma t } \,{\left({\mathrm{e}}^{\Gamma t } -1\right)}\,{\left(N_{B/F}+1\right)}}{{\left(2\,N_{B/F}+1\right)} } & 0 & 0\\
    0 & 0 & \frac{N_{B/F}-N_{B/F}\,{\mathrm{e}}^{-\Gamma t } }{{\left(2\,N_{B/F}+1\right)} } & 0\\
    {{\mathrm{e}}^{-\frac{\Gamma t - 2i \epsilon t}{2}} } & 0 & 0 & \frac{{\mathrm{e}}^{-\Gamma t } \,{\left(N_{B/F}+N_{B/F}\,{\mathrm{e}}^{\Gamma t } +1\right)}}{{\left(2\,N_{B/F}+1\right)} }
\end{array}\right)\\
\mathrm{}
\end{array}
\end{align}
{\it Remark.---}The only temperature dependence comes from $N_{B/F}$. In particular, the Hamiltonian term is independent of ($\theta$) and thus can be ignored. Then the optimal solution for this problem should be rotated with the same Hamiltonian in order to compensate for it. As such, we can ignore the phases in the off-diagonal terms above.

\subsection*{Using the expected mean logarithmic error as a cost function}
\begin{figure}
    \centering
    \includegraphics[width=0.8\linewidth]{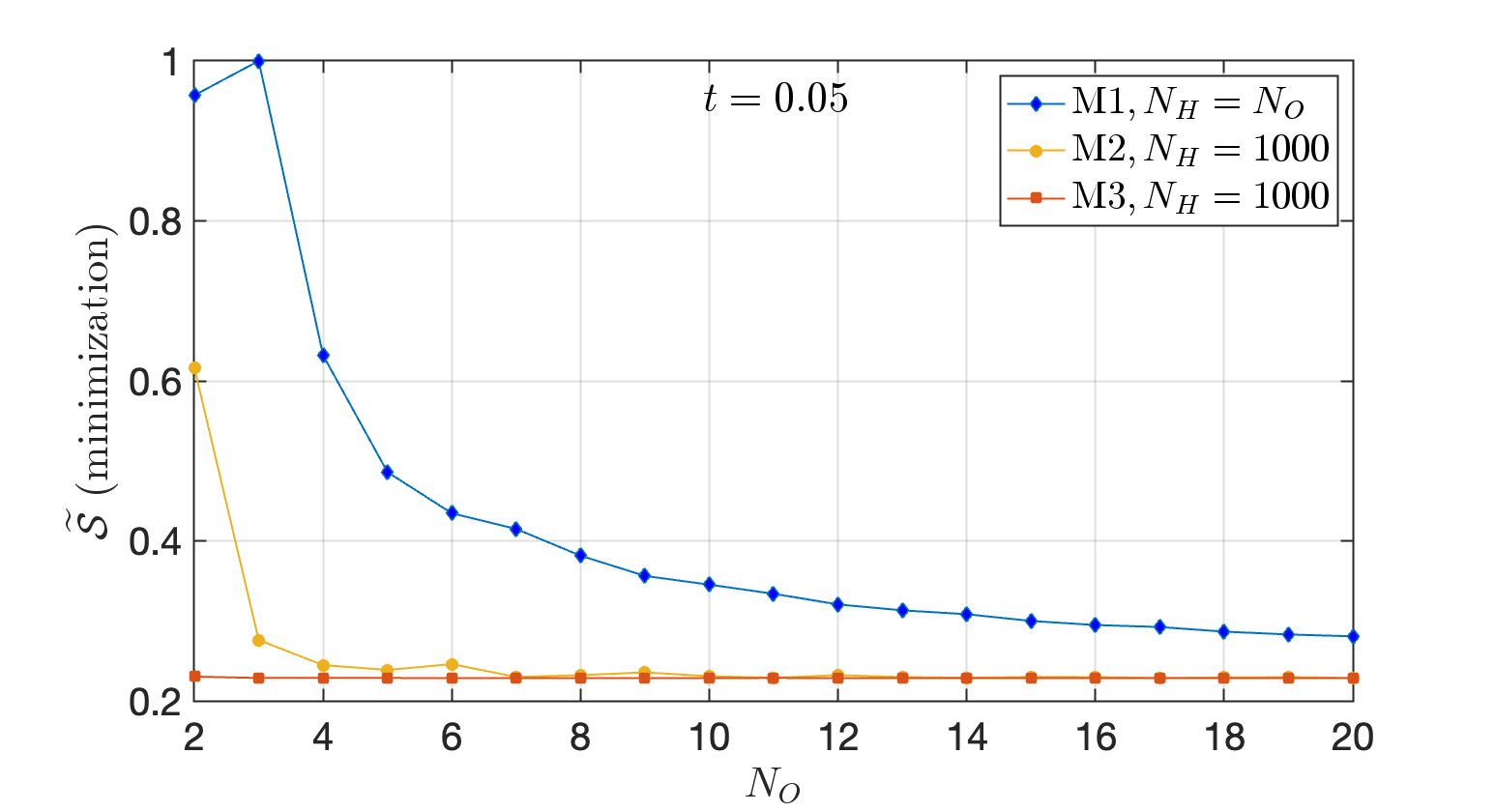}
    \includegraphics[width=0.8\linewidth]{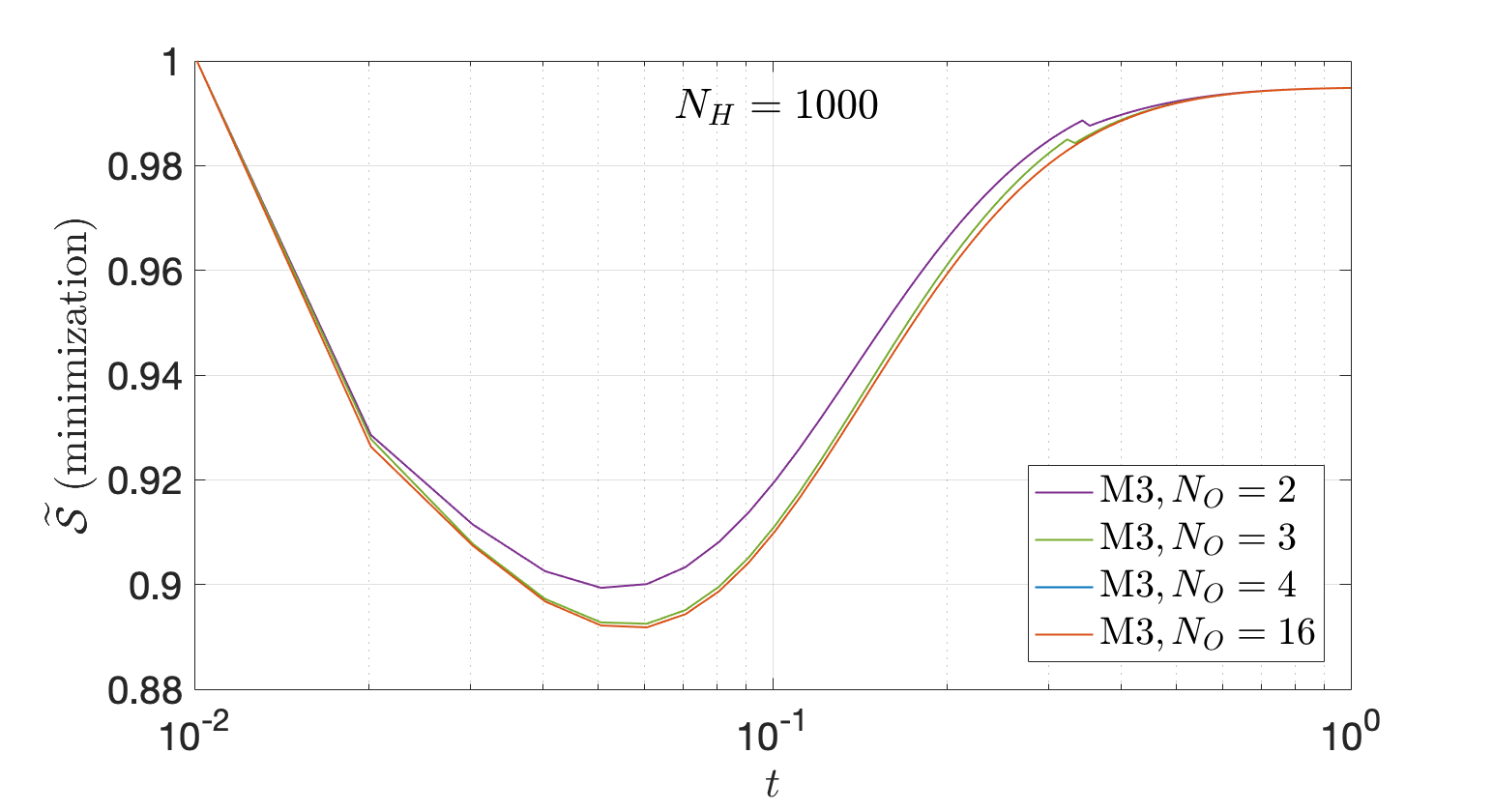}
    \includegraphics[width=0.8\linewidth]{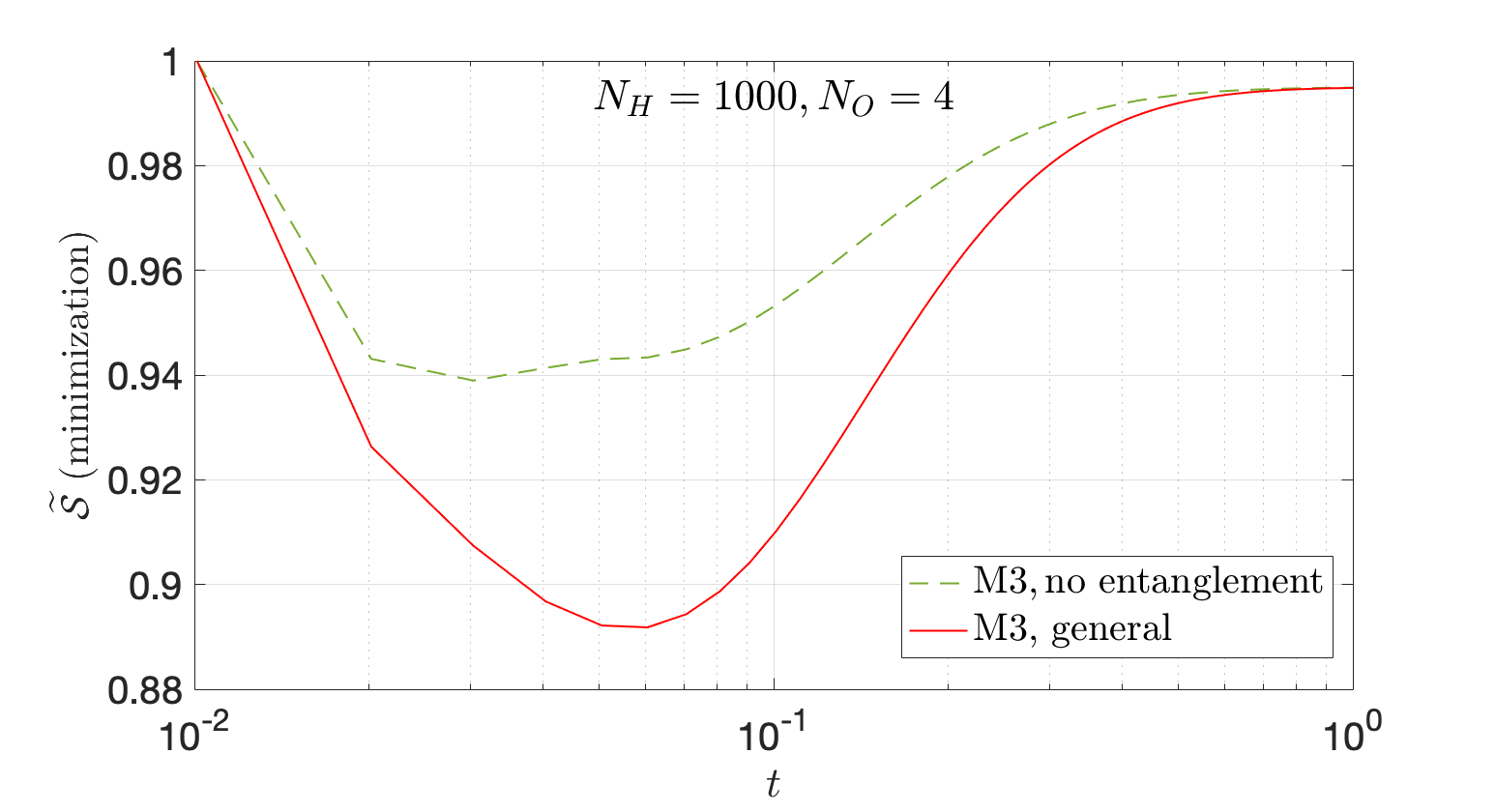}
    \caption{The thermometry problem seen from the perspective of the EMSLE as the cost function. The top, middle and low panels correspond to the Figs.~\ref{fig::allmethods_allNo_fixedtime},~\ref{fig::M3_differentNo_alltimes}, and~\ref{fig::noentanglement} of the main text, respectively---note the logarithmic scaling in the middle and bottom figures. All other parameters are kept the same as the corresponding graphs in the main text.}
    \label{fig:EMSLE}
\end{figure}

In the main text, we took the MSE as our figure of merit. However, in recent years, an alternative cost function has been put forward for thermometry, which is motivated by scale invariance~\cite{PhysRevLett.127.190402}. This is the so called expected mean square logarithmic error (EMSLE) at the kernel of which lies the following reward function
\begin{align}
r(\theta,\hat\theta^*_i) = \log^2(\hat \theta_i/\theta),
\end{align}
which can be analytically solved to find the optimal estimator as~\cite{PhysRevLett.127.190402}
\begin{align}
    \hat\theta^*_i = \exp{\int d\theta p(\theta|i) \log(\theta)}.
\end{align}
Interestingly, for this cost function, one can also prove that the optimal POVM is in fact a PVM~\cite{Rubio2022}. Our results straightforwardly apply to such figure of merit. We showcase this by reproducing our Figs.~\ref{fig::allmethods_allNo_fixedtime}, \ref{fig::M3_differentNo_alltimes}, and \ref{fig::noentanglement}. These are depicted here in the three panels of Fig.~\ref{fig:EMSLE}, respectively from top to bottom. The fact that PVMs are optimal is reflected in the middle panel, where our method M3 is optimal with only $N_O=4$ outcomes.

\bibliography{sensing}

\end{document}